\documentclass[conference]{IEEEtran}
\IEEEoverridecommandlockouts

\usepackage{cite}
\usepackage{amsmath,amssymb,amsfonts}
\usepackage{algorithmic}
\usepackage{graphicx}
\usepackage{booktabs} 
\usepackage{multirow}
\usepackage{textcomp}
\usepackage{xcolor}
\usepackage{subcaption}
\def\BibTeX{{\rm B\kern-.05em{\sc i\kern-.025em b}\kern-.08em
    T\kern-.1667em\lower.7ex\hbox{E}\kern-.125emX}}

\begin{document}

\title{AeroMesa: Efficient Data Management System for Multi-Dimensional Spatio-Temporal Trajectories
}

\author{
    \IEEEauthorblockN{
        Yue Zhang \textsuperscript{1}, 
        Zizhong Ding \textsuperscript{1}, 
        Lin Sun \textsuperscript{1}, 
        Haopeng Chen \textsuperscript{1,*},
        Yan Jiao \textsuperscript{2},
        Yongming Xu \textsuperscript{2}
    }
    \IEEEauthorblockA{
        \textsuperscript{1} Shanghai Jiao Tong University, Shanghai, China\\
        \textsuperscript{2} ShangHai Shapere Information Technology Co.,Ltd., Shanghai, China\\
    }
    \IEEEauthorblockA{
        Email: \{zhangyue20040611, enutrof65536, sun\_lin, chen-hp\}@sjtu.edu.cn;
        \{jiao, xu\}@shapere.xyz\\
    }
}

\maketitle

\begin{abstract}
The proliferation of multi-dimensional trajectory data---fueled by large-scale IoT and the emerging low-altitude economy, particularly UAV operations---drives repositories to jointly support $(x,y)$, $(x,y,t)$, $(x,y,z)$, and $(x,y,z,t)$ queries within a single storage framework. Yet existing HBase-based systems fall short in three respects: severe
row-key interval fragmentation when altitude is jointly encoded with horizontal coordinates, locality-unfriendly spatial encodings with workload-blind shape-code ordering, and coarse-grained temporal indexes that leave intra-slot boundary ambiguity unresolved. We present \textbf{AeroMesa}, an efficient data management system for multi-dimensional spatio-temporal trajectories built on Apache HBase and Redis, that natively supports $(x,y)$, $(x,y,t)$, $(x,y,z)$, and $(x,y,z,t)$ queries within a unified storage framework.
AeroMesa addresses the above limitations through three designs:
a decoupled horizontal-altitude architecture with a multi-granularity \textbf{Height Spatio-Temporal Index (HTSI)} that eliminates joint-encoding fragmentation; \textbf{Hilbert-BFS} with \textbf{Workload-Aware Jaccard (WAJ)} reordering that improves spatial locality; and \textbf{TI\textsuperscript{+}}, a dual-offset temporal index that resolves intra-slot false positives. Evaluations on T-Drive and a 87{,}537-trajectory high-fidelity UAV simulation demonstrate that AeroMesa reduces 3D/4D query latency by up to $30\times$ over XZ3/TXZ3, lowers 2D latency by up to 17.9\% over TMan, and cuts temporal candidates by up to 51.3\% over MCTM, with sub-linear scalability confirmed under $200\times$ data expansion, confirming AeroMesa's efficiency for multi-dimensional spatio-temporal trajectory management.
\end{abstract}

\begin{IEEEkeywords}
Trajectory Management, Spatio-Temporal Indexing
\end{IEEEkeywords}

\section{Introduction}

Trajectory data management has become a fundamental capability for intelligent transportation, logistics, and urban planning. Large-scale platforms continuously collect location traces from vehicles, couriers, and mobile sensing devices. For example, JD Logistics has reported that more than 60{,}000 couriers generate over 1\,TB of trajectory logs per day~\cite{ruan2020learning}. This trend is further intensified by the emerging low-altitude economy, where unmanned aerial vehicles (UAVs) are increasingly used for parcel delivery, infrastructure inspection, and emergency response. Compared with conventional ground trajectories, UAV traces introduce altitude as an extra dimension, driving trajectory repositories to jointly support $(x,y)$, $(x,y,t)$, $(x,y,z)$, and $(x,y,z,t)$ queries within a single storage framework.

Distributed key-value stores, particularly Apache HBase, are an attractive substrate for this setting because they offer horizontal scalability, high write throughput, and a mature operational ecosystem. Representative systems such as TrajMesa~\cite{li2021trajmesa}, VRE~\cite{lan2022vre}, and TMan~\cite{he2024tman} have shown that HBase can effectively support large-scale planar trajectory management. However, extending such systems from 2D workloads to unified multi-dimensional workloads is not as simple as lifting an existing 2D spatial encoding to a 3D or 4D one. A straightforward approach is to replace the 2D space-filling curve with a 3D one and jointly encode altitude with horizontal coordinates in a single spatial key space. In practice, this strategy is poorly matched to the anisotropic physical scales of trajectory dimensions: the horizontal domain usually spans large geographic regions, whereas UAV altitude is typically confined to a much narrower range. Once these dimensions are normalized into the same encoded space, practical query windows often become badly misaligned with the induced grid partitioning. In HBase, this misalignment manifests as a large number of disjoint row-key intervals, which increases RPC overhead and diminishes the benefit of sequential scans. Avoiding such row-key interval fragmentation while still supporting altitude predicates is therefore a primary challenge in HBase-based multi-dimensional trajectory management.

Beyond this architectural issue, the efficiency of existing spatial and temporal access paths is also limited. On the spatial side, TMan proposes a two-level approach: high-level XZ2 encodings for spatial cells and low-level geometric Jaccard reordering on intra-cell shape codes. Unfortunately, high-level XZ2 suffers from structural jumps inherent to Z-ordering which destroy physical locality. At the low level, geometric Jaccard reordering ignores the fact that, 
in skewed query-pattern workloads, specific shape-code patterns are more commonly co-accessed. On the temporal side, existing coarse-grained time-slot encodings often lose precise intra-slot boundary information, causing false positives. Taken together, these observations suggest that a unified HBase-based trajectory management system must solve two problems simultaneously: \textbf{it must avoid row-key interval fragmentation at the architectural level, and it should improve spatial locality and pruning precision at the access-path level.}

To this end, we present \textbf{AeroMesa}, a unified multi-dimensional trajectory management system built on a decoupled architecture that avoids forcing a highly anisotropic domain into a single joint spatial encoding. AeroMesa utilizes a primary \textbf{Spatial Index (SI)} that encodes only $xy$ (longitude-latitude) information, combined with a server-side \textbf{ZFilter} which pushes down altitude predicates, to continuously preserve horizontal locality while enabling efficient altitude pruning for 3D queries. For 4D spatio-temporal queries, we introduce a secondary index table, called the \textbf{Height Spatio-Temporal Index (HTSI)}, which leverages multi-granularity altitude slots to prune away data except the queried altitude slot. This confines the scan to only the relevant altitude band rather than sweeping across all heights.


Within this decoupled architecture, AeroMesa further enhances the underlying access paths. To improve spatial locality, AeroMesa proposes a \textbf{Hilbert-BFS} encoding coupled with a \textbf{Workload-Aware Jaccard (WAJ)} reordering metric. Unlike purely geometric Jaccard, WAJ blends shape-code similarity with query-workload co-access frequency. For temporal precision, AeroMesa introduces \textbf{TI+}, an enhanced temporal index that tracks fine-grained intra-bucket boundaries. By supporting exact intra-bucket boundary checks at the key-value level, TI+ can reduce false positives in temporal range queries and lower overall query latency.

We implement AeroMesa on top of Apache HBase and Redis. Extensive evaluations are conducted using a large-scale trajectory dataset (T-Drive) and a high-fidelity UAV dataset, the latter containing approximately 90{,}000 simulated flights over a $10\,\mathrm{km}\times10\,\mathrm{km}$ low-altitude airspace in Shanghai.

The main contributions of this paper are as follows:
\begin{itemize}
    \item We design an efficient multi-dimensional 
    spatio-temporal trajectory data management system.
    The system natively supports \((x,y)\), \((x,y,t)\), \((x,y,z)\), 
    and \((x,y,z,t)\) queries within a unified storage framework. 
    By eliminating row-key interval fragmentation through a decoupled 
    horizontal-altitude architecture, it reduces query latency by \(10\times\)--\(30\times\) 
    over XZ3 and TXZ3 joint encodings for 3D and 4D workloads, 
    with contiguous scan intervals reduced by up to three orders of magnitude. 
    For 4D queries, HTSI further reduces latency by 21\%--54\% 
    and cuts the candidate segment set by up to 76\%.

    \item We propose a suite of novel indexing techniques.
    Hilbert-BFS encoding, coupled with Workload-Aware Jaccard
    reordering, improves planar spatial locality, reducing latency by 
    8.3\%--17.9\% over TMan and 84.4\%--89.3\% over GeoMesa on 2D queries.
    TI\textsuperscript{+} augments time-slot encoding with exact 
    intra-bucket boundary checks, reducing candidate segments by up to 
    51.3\% and average temporal query time by up to 21.2\% over MCTM~\cite{tao2025mctm} for short temporal windows.

    \item We implement AeroMesa on Apache HBase and Redis, demonstrating robust scalability across both read and write operations. Under a 200$\times$ data-volume expansion, query latency increases by only 11.03$\times$, confirming sub-linear read scalability. Meanwhile, amortized write costs scale gracefully, showing only a 4.3$\times$ increase against a 1000$\times$ growth in store size, proving its scalable efficiency in handling large-scale, high-density trajectory data.
    
\end{itemize}
\section{Related Work}
\label{sec:related}

\subsection{Trajectory Management Systems}

Early single-machine systems such as TrajStore~\cite{cudre2010trajstore} and Torch~\cite{wang2018torch} suffer from inherent scalability limits. Distributed and in-memory systems---including ST-Hadoop~\cite{alarabi2018st}, Summit~\cite{alarabi2018summit}, UlTraMan~\cite{ding2018ultraman}, DITA~\cite{shang2018dita}, TrajSpark~\cite{zhang2017trajspark} and Dragoon~\cite{fang2021dragoon}---improve throughput and scale but remain uneconomical for large-scale historical trajectory storage or support only a limited query scope. Relational extensions such as MobilityDB~\cite{zimanyi2020mobilitydb} and spatial Spark frameworks such as Simba~\cite{xie2016simba} and GeoSpark~\cite{yu2015geospark} offer complementary capabilities but are either limited in scalability or provide insufficient spatio-temporal support.

Modern NoSQL-based systems differ primarily in storage granularity. Point-based systems (e.g., GeoMesa~\cite{hughes2015geomesa}) support precise queries but suffer from severe write amplification. Trajectory-based systems (e.g., TrajMesa, TMan) avoid write amplification but must scan entire trajectories to answer range queries. Segment-based systems (e.g., VRE, MCTM) offer a practical middle ground. On the indexing side, early systems such as TrajMesa and THBase~\cite{qin2019thbase} replicate data across multiple primary tables, incurring high storage costs; later systems such as TMan, VRE, and MCTM replace this with a single primary table plus lightweight secondary indexes.

However, no existing system provides unified support for $(x,y)$, $(x,y,t)$, $(x,y,z)$, and $(x,y,z,t)$ queries within a single framework---a gap AeroMesa fills for UAV and low-altitude trajectory workloads.

\subsection{Spatial Index}

R-tree~\cite{guttman1984r} and its variants~\cite{beckmann1990r,tao2003tpr} are widely used for multi-dimensional indexing, but their dynamic rebalancing incurs significant write latency under high-throughput ingestion, making them unsuitable for large-scale HBase-backed systems~\cite{yu2016two}. Hilbert R-tree~\cite{kamel1993hilbert} and CSE-tree~\cite{wang2008flexible} improve clustering but retain dynamic tree structures with prohibitive maintenance overhead in write-intensive key-value stores.

Geohash encodes 2D coordinates via Z-order curves but does not extend to range-bearing objects such as trajectory segments. XZ-Ordering~\cite{boxhm1999xz} addresses this with double-enlarged elements representing segment MBRs, and has been widely adopted by GeoMesa, TrajMesa, JUST~\cite{li2020just}, VRE, TraSS~\cite{he2022trass}, MCTM, and TMan. Subsequent work refined the approach: TrajMesa introduced XZ2+ with PosCode, MCTM proposed an Index Cache, and TMan appends Jaccard-ordered shape codes to reduce scan intervals.

Hilbert curves offer better locality than Z-order curves~\cite{moon2001analysis} and have been studied for point data~\cite{liu2023hgst}, but no prior work has proposed a Hilbert-based enlarged-element strategy for trajectory segments. For 3D indexing, naively extending XZ2 to XZ3 treats altitude as an equivalent spatial dimension, causing severe resolution imbalance for UAV trajectories whose vertical range is orders of magnitude smaller than their horizontal extent. AeroMesa instead encodes only horizontal coordinates via Hilbert-BFS and handles altitude filtering independently through HBase-side ZFilters.

\subsection{Temporal Index}

Interval trees and segment trees support exact interval matching but are ill-suited for key-value databases under concurrent ingestion. Most trajectory systems therefore adopt a time-binning strategy. ST-Hadoop establishes this paradigm but produces false positives when query windows are narrower than a bin. TrajMesa's XZT index introduces up to $1/2$ temporal dead space per level; VRE indexes by start time bin, causing wide over-scans; TMan records start and end bins at coarse granularity, materializing all boundary-intersecting segments. MCTM improves upon these with a single intra-SegTime offset, but for short segments fully inside one slot, the unknown opposite boundary still introduces false positives.

\subsection{Spatio-Temporal Index}

TB-tree~\cite{pfoser2000novel} bundles spatiotemporally adjacent segments for improved locality but scales poorly due to dynamic tree maintenance. 3D R-tree~\cite{zhu2007efficient} treats time as a third spatial dimension, but bounding-box overlap grows substantially along the time axis for long-term data. JUST-Traj~\cite{he2021just-traj} partitions time into intervals with separate spatial indexes per period, but uneven grid-level distribution fragments queries into many discrete scan intervals. MCTM structures its TSI as \textit{BinNum} + \textit{SI} + \textit{SegTimeID}, improving I/O efficiency on key-value stores.

None of these systems index the altitude dimension, and folding altitude into the spatial encoding via XZ3 introduces severe scale imbalance.
\section{Framework}
\label{sec:framework}
\begin{figure}[h]
    \centering
    \includegraphics[width=1.02\linewidth]{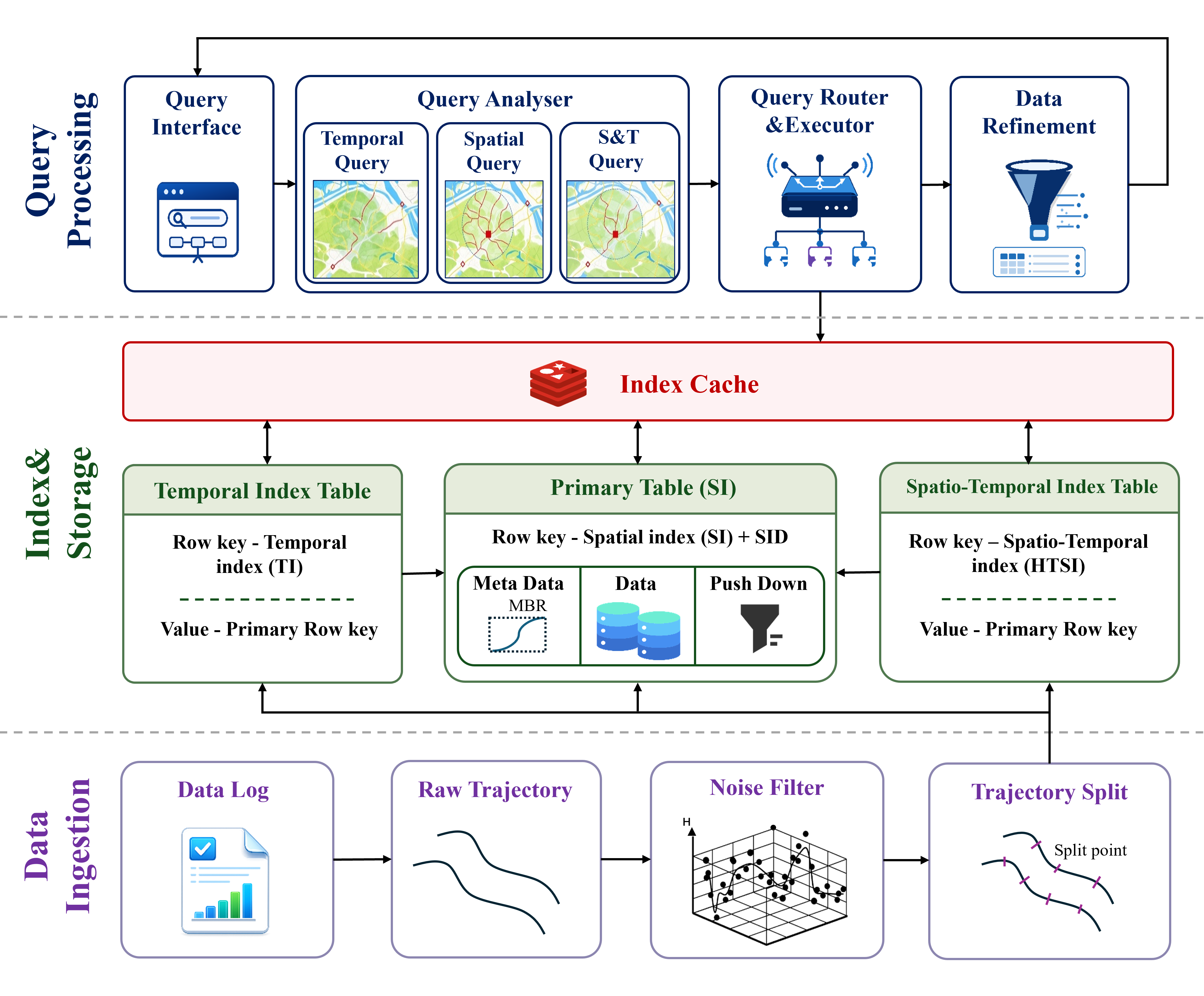} 
    \caption{The Framework of AeroMesa}
    \label{fig:framework}
\end{figure}

Figure~\ref{fig:framework} illustrates the overall architecture of AeroMesa, which consists of three main modules: Data Ingestion, Indexing \& Storage, and Query Processing. 

\textbf{Data Ingestion.} To achieve high-throughput trajectory data ingestion, this module implements a three-stage pipeline consisting of: 1) memory-mapped parsing, which utilizes Memory-Mapped I/O to map large files directly into off-heap memory, avoiding user-space data copying; 2) outlier filtering, which eliminates erroneous data points caused by signal instability; and 3) index-aligned segmentation, which dynamically partitions the continuous trajectory logs into atomic segments based on the pre-defined boundaries.

\textbf{Indexing \& Storage.}
To support diverse multi-dimensional queries within a unified system, AeroMesa adopts a primary-secondary table architecture. After segmentation, it organizes all data around a single primary storage table and persists the full system state across Apache HBase and Redis.

The primary table (SI table) uses a composite row key of the spatial index key and segment ID (SID). Each row stores the raw point sequence as a binary blob together with the trajectory MBR metadata, which serve as lightweight server-side attributes for pushdown pre-filtering. On top of this foundation, AeroMesa maintains three novel index structures.

\begin{figure*}[t!] 
\clearpage 
  \centering
  \includegraphics[width=\textwidth]{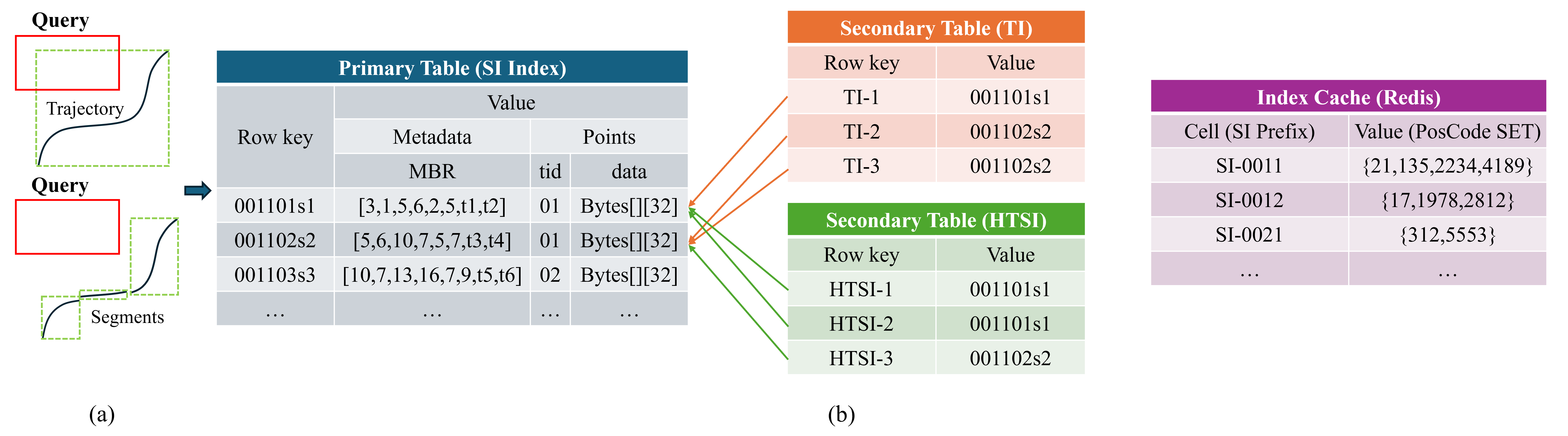}
  \caption{Overview of AeroMesa's three-tier storage schema: 
           primary segment table, secondary index tables, and Index Cache.}
  \label{fig:storage-schema}
\end{figure*}

The secondary index tables are backed by HBase by default, following the design of TMan. AeroMesa additionally provides a Redis-backed variant for both secondary indexes, allowing operators to trade storage cost for lower scan latency. Furthermore, AeroMesa maintains a per-cell shape-code set in Redis as a cache layer, enabling rapid in-memory shape-code pruning and substantially reducing false-positive candidates at negligible latency cost.

\textbf{Query Processing.}
To efficiently support temporal, 2D/3D spatial, and 4D spatio-temporal queries, \textsc{AeroMesa} coordinates a three-stage processing pipeline. Each query is first parsed by a Query Analyser to identify its target dimensions, and then forwarded to a lightweight Query Router, which dispatches the query execution plan to the appropriate indexing layer for accelerated retrieval.

For \textit{spatial queries}, the executor issues range scans directly 
against the primary SI table. Altitude predicates are pushed down 
to the HBase region server via a ZFilter, so mismatched segments 
are directly skipped before data leaves the server.

For \textit{temporal queries} and \textit{spatio-temporal queries}, the executor scans the secondary index for candidate IDs, then fetches payloads from the primary SI table with server-side filters (temporal boundary and ZFilter) applied.

Regardless of routing path, all queries conclude with a \textit{refinement phase} that applies exact geometric verification on the retrieved payloads and returns the final results.

\section{Indexing and Storing}
\label{sec:index}
This section presents the indexing and storage design of AeroMesa.
We begin with the trajectory data model that governs how raw point
sequences are partitioned and stored (Section~\ref{sec:model}), then
describe our storage schema built atop this model
(Section~\ref{sec:storage}). With the storage layout established, we
introduce two index structures that accelerate query resolution: a
two-level spatial index combining Hilbert-BFS macroscopic encoding
with Workload-Aware Jaccard microscopic reordering
(Section~\ref{sec:spatial}), and a temporal index
TI\textsuperscript{+} that eliminates false-positive fetches for
short intra-interval trajectories (Section~\ref{sec:ti}). We then
extend this foundation to the anisotropic dimensional characteristics
of 3D/4D trajectory data via a server-side ZFilter that prunes
altitude-mismatched segments without joint encoding
(Section~\ref{sec:zfilter}) and the HTSI Index
(Section~\ref{sec:htsi}).

\subsection{Trajectory Model}
\label{sec:model}

A trajectory is generated by a moving object reporting its position
at regular intervals. Each \emph{trajectory point} is a tuple
$p = \langle x, y, z, t \rangle$, where $x$ and $y$ denote horizontal
coordinates, $z$ denotes altitude, and $t$ denotes the timestamp.
A full trajectory $\mathit{Traj}$ is an ordered sequence of points
$P = \{p_0, p_1, \ldots, p_n\}$ sorted by timestamp.

Storing a full trajectory as a single row (trajectory-based model)
produces large MBRs that poorly approximate the actual spatial extent,
leading to false positives in range queries. Storing each
point as an independent row (point-based model) incurs high write
amplification and per-point indexing overhead.
Following the segment-based paradigm of VRE and MCTM, as represented
in Figure~\ref{fig:storage-schema}(a), AeroMesa partitions each
trajectory into a sequence of non-overlapping segments.

\begin{figure*}[t!]
  \centering
  \includegraphics[width=\textwidth]{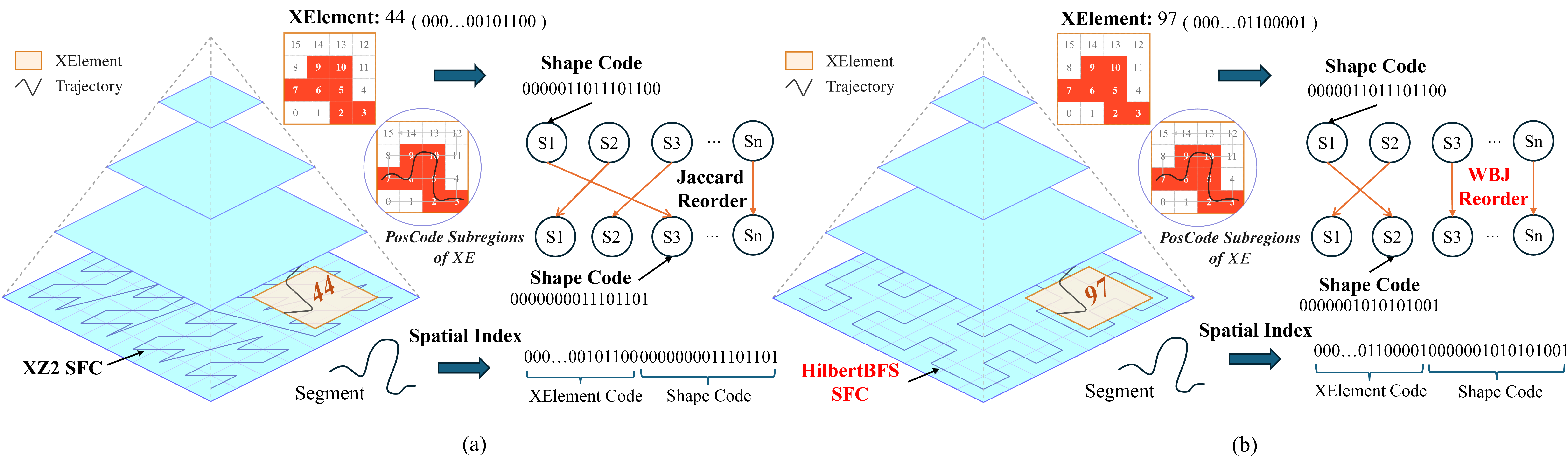}
  \caption{Spatial index encoding pipeline: (a)~TMan; (b)~AeroMesa.}
  \label{fig:si-encoding}
\end{figure*}

AeroMesa uses fixed-duration time windows to partition trajectories:
two consecutive points $p_i$ and $p_{i+1}$ are assigned to different
segments when they cross a window boundary, formally defined as
$\lfloor p_i.t \,/\, \mathit{dur} \rfloor \neq
\lfloor p_{i+1}.t \,/\, \mathit{dur} \rfloor$,
where $\mathit{dur}$ is the fixed window duration. This yields a
sequence of non-overlapping segments
$\mathit{Segs} = \{\mathit{seg}_0, \mathit{seg}_1, \ldots,
\mathit{seg}_m\}$, each stored as a single HBase row.

\subsection{Storage Schema}
\label{sec:storage}

AeroMesa organises all persistent state across three tiers, as shown
in Figure~\ref{fig:storage-schema}(b):

\subsubsection*{Primary Table}
The primary table stores the segment data together with its
multidimensional metadata. AeroMesa constructs the row key by
concatenating three fields:

\begin{equation}
  \mathit{rowkey} =
  \underbrace{\text{\small\texttt{partition\_id}}}_{\text{configurable prefix}}
  \;\text{``-''}\;
  \underbrace{\mathit{SI}}_{\text{Spatial Index}}
  \;\text{``-''}\;
  \underbrace{\mathit{sid}}_{\text{segment id}}
  \label{eq:rowkey}
\end{equation}

The \texttt{partition\_id} prefix controls the trade-off
between locality and load balance and can be tuned per
deployment, for example via equal-frequency quantiles of
the rowkey distribution to ensure balanced region load in
distributed settings. The packed spatial index
$\mathit{SI}$ (as detailed in Section~\ref{sec:spatial}) ensures
that spatially adjacent segments cluster within each partition. The segment identifier $\mathit{sid}$ uniquely identifies each segment.

Each row stores two parts. The \textbf{metadata} column
family holds the segment's four-dimensional MBR, which are
evaluated by server-side filters \emph{before} the payload is read,
enabling spatial, altitude, and temporal pruning without decoding raw
points. The \textbf{payload} column family stores
the serialised point list as a compact binary blob ($32\,\text{bytes}$ per
point).

\subsubsection*{Secondary Tables}
To support multiple query types without duplicating raw payloads,
AeroMesa maintains lightweight secondary tables that map index keys
(e.g., HTSI or TI\textsuperscript{+} codes) to the corresponding
primary row keys. These tables can be backed
by either Redis or an HBase table depending on deployment requirements.

\subsubsection*{Index Cache}
Verifying all shape codes in a spatial cell against the query window on every query incurs unnecessary I/O overhead. AeroMesa addresses this by maintaining a Redis SET per spatial cell that records every distinct shape code ever written into that cell, following the design of TMan and MCTM. At query time, a single set-membership
lookup retrieves the populated codes; intersecting this set with the
geometrically overlapping shape codes for the query window yields a small
candidate list, each entry then resolved via a targeted score-range scan.
To bound memory consumption, the cache is capped at a configurable maximum of $N$ entries. When the cap is reached, the least-recently-used entry is evicted. Eviction introduces only a cache miss on the subsequent access, which falls back to the downstream scan---preserving correctness.

\subsection{Spatial Index}
\label{sec:spatial}

The storage schema above places segments into HBase rows ordered by
a 1D key. The central challenge of spatial indexing is to
assign 1D keys such that segments close in 2D space remain close on
the key axis, minimising the number of disjoint HBase range scans
required to answer a spatial query.

AeroMesa builds on the two-level spatial encoding paradigm of
TMan's TShape.
In this paradigm, a \emph{macroscopic} space-filling
curve (SFC) maps each segment to a primary grid cell, while a
\emph{microscopic} $4\!\times\!4$ bitmask shape code
$s \in \{1,\ldots,2^{16}-1\}$ captures the intra-cell footprint of
the segment. TShape uses a Z-order SFC at the macroscopic level and
a geometric Jaccard similarity shape-code reordering at the microscopic
level (Figure~\ref{fig:si-encoding}(a)). Both choices carry known weaknesses: Z-order scatters spatially
adjacent cells across discontinuous key intervals, and treating all
sub-cell patterns equally wastes disk locality on rarely queried
shapes. AeroMesa renovates both layers with \textbf{Hilbert-BFS}
(where BFS denotes a breadth-first traversal of the quad-tree,
ensuring that cells at the same resolution level remain contiguous
in the encoding) and \textbf{Workload-Aware Jaccard
(WAJ)}, yielding the pipeline shown in Figure~\ref{fig:si-encoding}(b).

\subsubsection{Macroscopic Encoding: Hilbert-BFS}
\label{sec:hbfs}

AeroMesa retains the enlarged-cell construction and per-axis binary
bisection of the TShape quad-tree, but replaces Z-order traversal
with a 2D Hilbert-curve ordering. The Hilbert index of a segment is:
\begin{equation}
  \label{eq:hbfs}
  \mathit{Index}_{H\text{-}\mathrm{BFS}}(\mathit{seg})
  =\sum_{k=1}^{g-1}4^k+\mathcal{H}_{2D}(C_i),
\end{equation}
where $g$ is the quad-tree depth at which the enlarged cell $E(C_i)$
first contains $\mathit{MBR}(\mathit{seg})$, found by a BFS over the
quad-tree, and $\mathcal{H}_{2D}(\cdot)$ is the standard 2D Hilbert
index of the host cell $C_i$. The offset $\sum_{k=1}^{g-1}4^k$ ensures
that cells at different depths map to disjoint key ranges, preserving
injectivity across levels. A correct BFS termination requires that
every MBR has a well-defined host cell; this is guaranteed by the
enlargement strategy: at level $l$ with cell side length $\delta_l$,
the base cell is extended by $2\delta_l$ in the positive $x$ and $y$
directions, so any MBR whose width and height are both smaller than
$\delta_l$ is guaranteed to fall within the enlarged cell. Ties among
siblings are broken deterministically by minimum Hilbert index.

\paragraph{Why Hilbert outperforms Z-order?}
The benefit of the Hilbert curve is a reduction in the
number of disjoint HBase \texttt{Scan} calls required to cover a
query window. Moon et al.\ prove the asymptotic
expectations for a $2^k\!\times\!2^k$ query region:
\begin{equation}
  \label{eq:cluster-hilbert}
  \mathbb{E}\!\bigl[\#\mathrm{clusters}_{\mathrm{Hilbert}}\bigr]
  \;\sim\;\frac{2^k}{3},
  \qquad
  \mathbb{E}\!\bigl[\#\mathrm{clusters}_{\mathrm{Z}}\bigr]
  \;\sim\;\frac{2^k}{2},
\end{equation}
as $k\to\infty$ -- a 33\% asymptotic reduction in scan
fragments.

\smallskip
\noindent\emph{Empirical confirmation.}
On the T-Drive dataset~\cite{yuan2010t}, holding shape code bits
constant to isolate the macroscopic layer, Hilbert-BFS reduces HBase scan intervals by 40\% versus XZ2.

\subsubsection{Microscopic Reordering: Workload-Aware Jaccard (WAJ)}
\label{sec:WAJ}

\begin{figure}[htbp]
  \centering
  \includegraphics[width=0.94\linewidth]{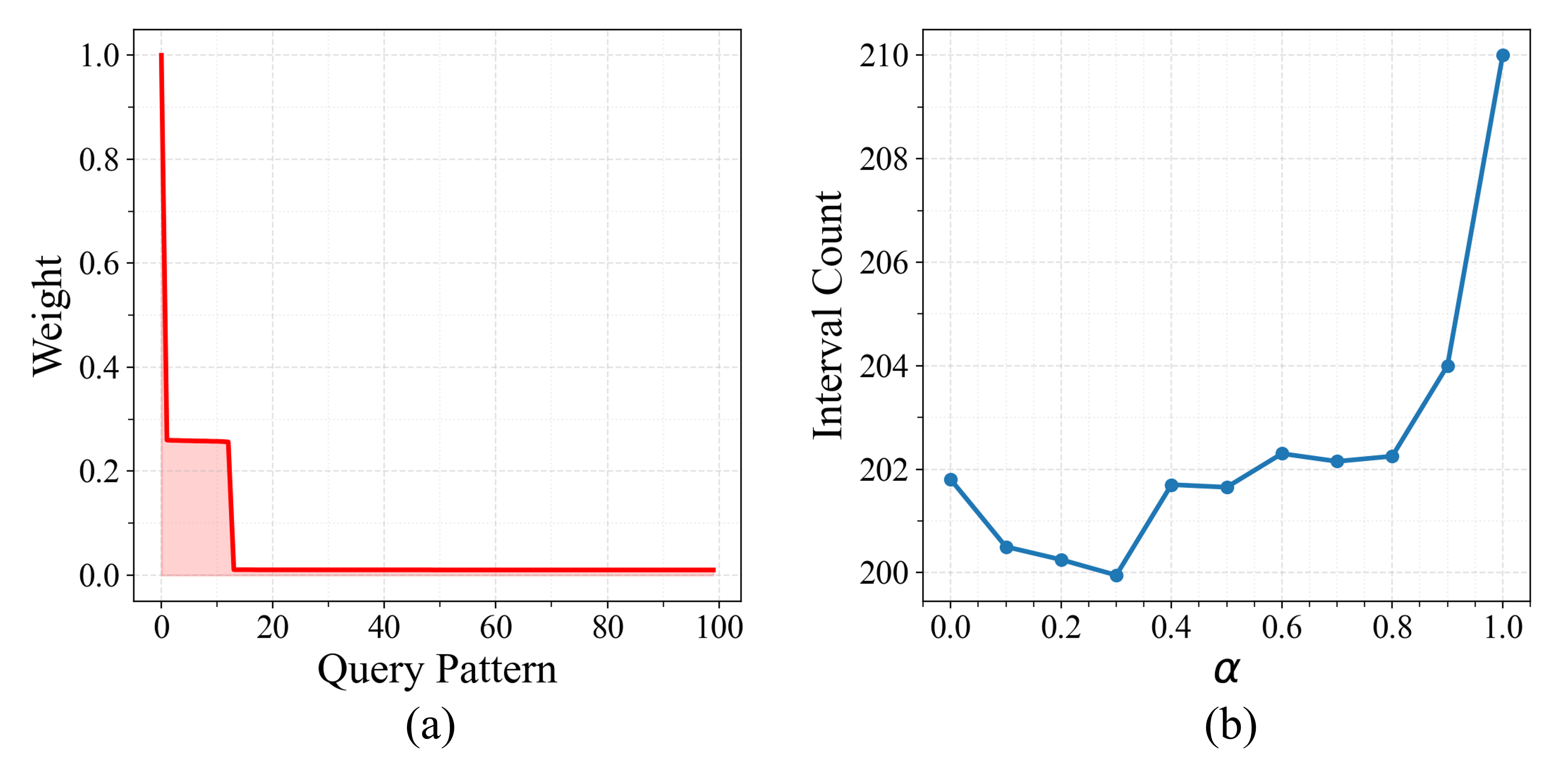}
  \caption{WAJ workload analysis: (a)~estimated pattern weight $\hat{P}(k)$ (sorted by weight); (b)~$\alpha$ ablation (square workload).}
  \label{fig:WAJ-reorder}
\end{figure}

The Hilbert-BFS encoding determines which HBase \emph{cell} a segment
falls into; within that cell, the shape code determines which
\emph{sub-range} of the cell is scanned. TMan's original ordering
relies on geometric Jaccard similarity, treating all query patterns
as equally likely. In practice, however, query workloads are skewed:
a handful of boundary-crossing query patterns dominate query frequency. An ordering blind to this skew fails to
cluster frequently co-accessed patterns into contiguous sub-ranges,
fragmenting scans across scattered regions. AeroMesa addresses this
with Workload-Aware Jaccard, which reorders shape codes according to
observed query pattern frequency, in three steps.

\paragraph{Step 1: Bounding the achievable pattern space}
Any axis-aligned query window $Q$ intersects a cell in a
\emph{contiguous rectangular block} $[r_1,r_2]\times[c_1,c_2]$ of
sub-cells. Since a $4\!\times\!4$ grid has 5 possible split positions
per axis, the set $\mathcal{K}$ of patterns that can actually be
activated by a query satisfies:
\begin{equation}
  \label{eq:K-bound}
  |\mathcal{K}|\;\le\;\tbinom{5}{2}^{\!2}=100,
\end{equation}
a $655\times$ reduction from the naïve $2^{16}$ upper bound, so the
prior need only be estimated over 100 patterns rather than $2^{16}$.

\paragraph{Step 2: Estimating the query pattern frequency}
For each achievable query pattern $k\in\mathcal{K}$, let $P(k)$ be the
probability that a random boundary-intersecting cell exhibits pattern
$k$ under the observed workload. We estimate $P(k)$ via offline
Monte Carlo sampling over $N=100{,}000$ i.i.d.\ queries:
\begin{equation}
  \hat{P}(k) = \frac{\text{boundary cells exhibiting pattern }k}
                    {\text{total boundary cells across all queries}}.
  \label{eq:Pk}
\end{equation}
By the law of large numbers, $\hat{P}(k)$ converges to $P(k)$ at rate
$O(N^{-1/2})$, so $N=100{,}000$ samples yield stable estimates.
The resulting distribution is heavily skewed
(Figure~\ref{fig:WAJ-reorder}(a)): fewer than 12 of the 100 achievable
patterns carry substantial weight, confirming the co-access
concentration that a workload-aware ordering can exploit.

\paragraph{Step 3: The WAJ similarity metric}
Given $\hat{P}(k)$, the \textbf{CoHit} score measures how often two
shape codes $A$ and $B$ are simultaneously activated by the same
query pattern, where $A\,\&\,k\neq 0$ indicates that shape code $A$
is activated by query pattern $k$:
\begin{equation}
  \label{eq:cohit}
  \mathit{CoHit}(A,B)
  =\sum_{k\in\mathcal{K}}\hat{P}(k)\cdot
   \mathbf{1}[A\,\&\,k\neq 0]\cdot\mathbf{1}[B\,\&\,k\neq 0].
\end{equation}
The final WAJ similarity combines CoHit with the geometric Jaccard
from TShape to preserve coarse spatial structure:
\begin{equation}
  \label{eq:WAJ}
  \operatorname{WAJ}(A,B) = \alpha \cdot \operatorname{\widetilde{CoHit}}(A,B)
  + (1-\alpha) \cdot \operatorname{\widetilde{Jacc}}(A,B),
\end{equation}
where $\widetilde{\,\cdot\,}$ denotes min-max normalisation over all
observed pairs. The mixing weight $\alpha=0.30$ is selected by
ablation over $\alpha\in\{0.1,0.2,\ldots,0.9\}$: the interval count is
minimized near $\alpha\approx0.30$ and stays flat over
$\alpha\in[0.2,0.5]$ (Figure~\ref{fig:WAJ-reorder}(b)), so the result
is insensitive to the exact weight. Shape codes are then linearised by solving a maximum-weight Hamiltonian path on the WAJ similarity graph; following TMan, we approximate this NP-hard problem with a genetic algorithm. The resulting ordering is computed once at write time, remaining transparent to the query path.

\paragraph{Robustness to workload drift}

\begin{figure}[htbp]
  \centering
  \includegraphics[width=0.94\linewidth]{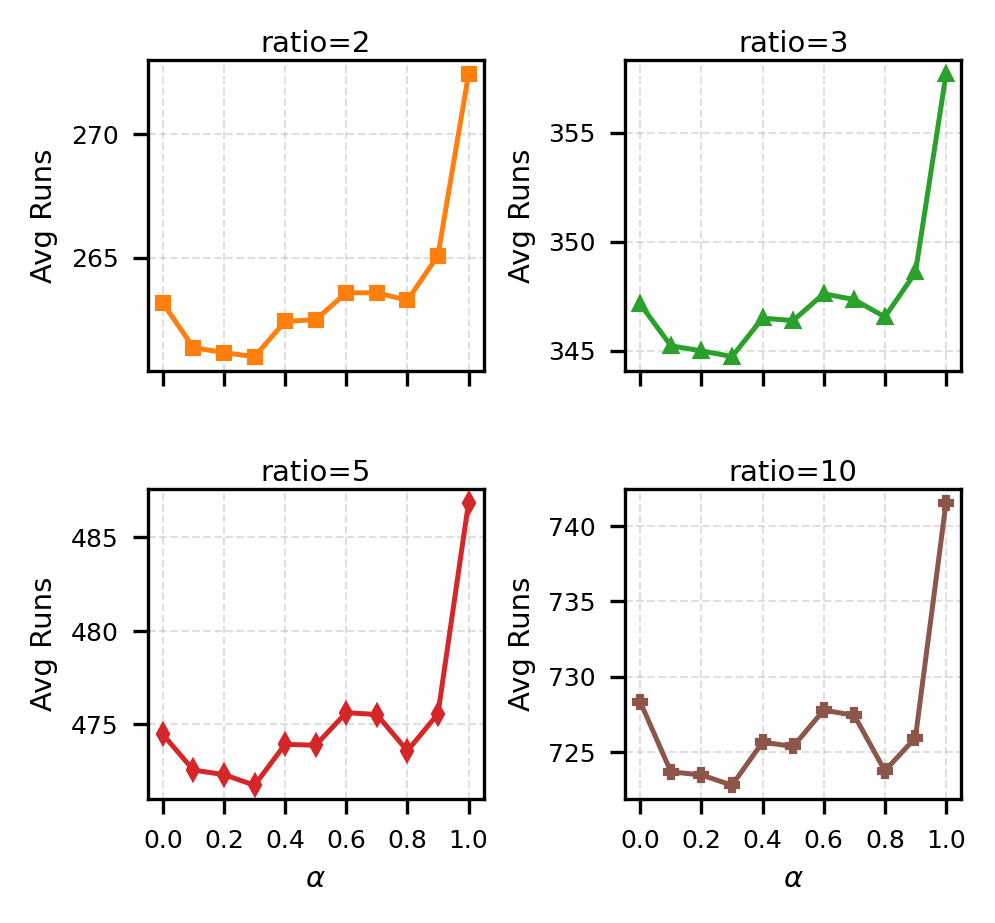}
  \caption{WAJ cross-distribution robustness}
  \label{fig:WAJ-sensitive-ratio}
\end{figure}

By definition, $\hat{P}(k)$ depends strictly on \emph{how} a window intersects a cell, guaranteeing invariance to query-center and absolute-size drift by construction. We further stress-test the sole remaining drift axis: aspect ratio. All curves in Figure~\ref{fig:WAJ-sensitive-ratio} use a single weight vector $\hat{P}(k)$ estimated under a square workload, evaluated on progressively more elongated queries (up to $10\!:\!1$ aspect ratio). Notably, across this entire range, the optimal $\alpha$ remains near 0.30 and the interval count stays low and flat. Collectively, these invariance properties imply that a single offline-estimated $\hat{P}(k)$ remains highly robust across the full space of workload shifts and thus requires no runtime re-tuning.

\subsubsection{Update}
\label{sec:update}
\noindent On the update path, we introduce a flat main+delta LSM
structure. Incoming segments are appended only to a lightweight
delta index, avoiding per-write global WAJ re-encoding of the main
index. Once the delta reaches a configurable threshold, a background
compaction merges it into the main index in batch while queries
continue to access the current version, guaranteeing read/write
consistency throughout. This design decouples the expensive global
reordering from the synchronous write path, reducing amortized write
latency.

\subsection{Temporal Index: TI\textsuperscript{+}}
\label{sec:ti}

The spatial index improves locality to minimise disjoint range scans; the temporal index then prunes candidates that fall outside the query range. AeroMesa inherits the base
temporal index structure from MCTM, which partitions the
timeline into day bins and hourly intervals, assigning each segment
a 32-bit packed index encoding day-bin, hourly-interval number, segment
type, and a 9-bit intra-interval offset ($\approx\!7\,\text{s}$).
MCTM defines three segment types: \emph{type 2} (segment starts within
this interval), \emph{type 1} (segment spans the entire interval),
and \emph{type 0} (segment terminates within this interval).

\paragraph{TI\textsuperscript{+}: dual-offset type 3}
\label{subsec:ti-dual}
A segment whose start and end both fall within the \emph{same}
hour interval is classified by MCTM as type 2, discarding the
end-time information. At query time, the system cannot determine
whether such a segment overlaps $[t_s, t_e]$ without fetching its
raw points from HBase, forcing an unnecessary HBase fetch for
segments that lie entirely outside $[t_s, t_e]$.

\begin{figure}[htbp]
  \centering
  \includegraphics[width=1\linewidth]{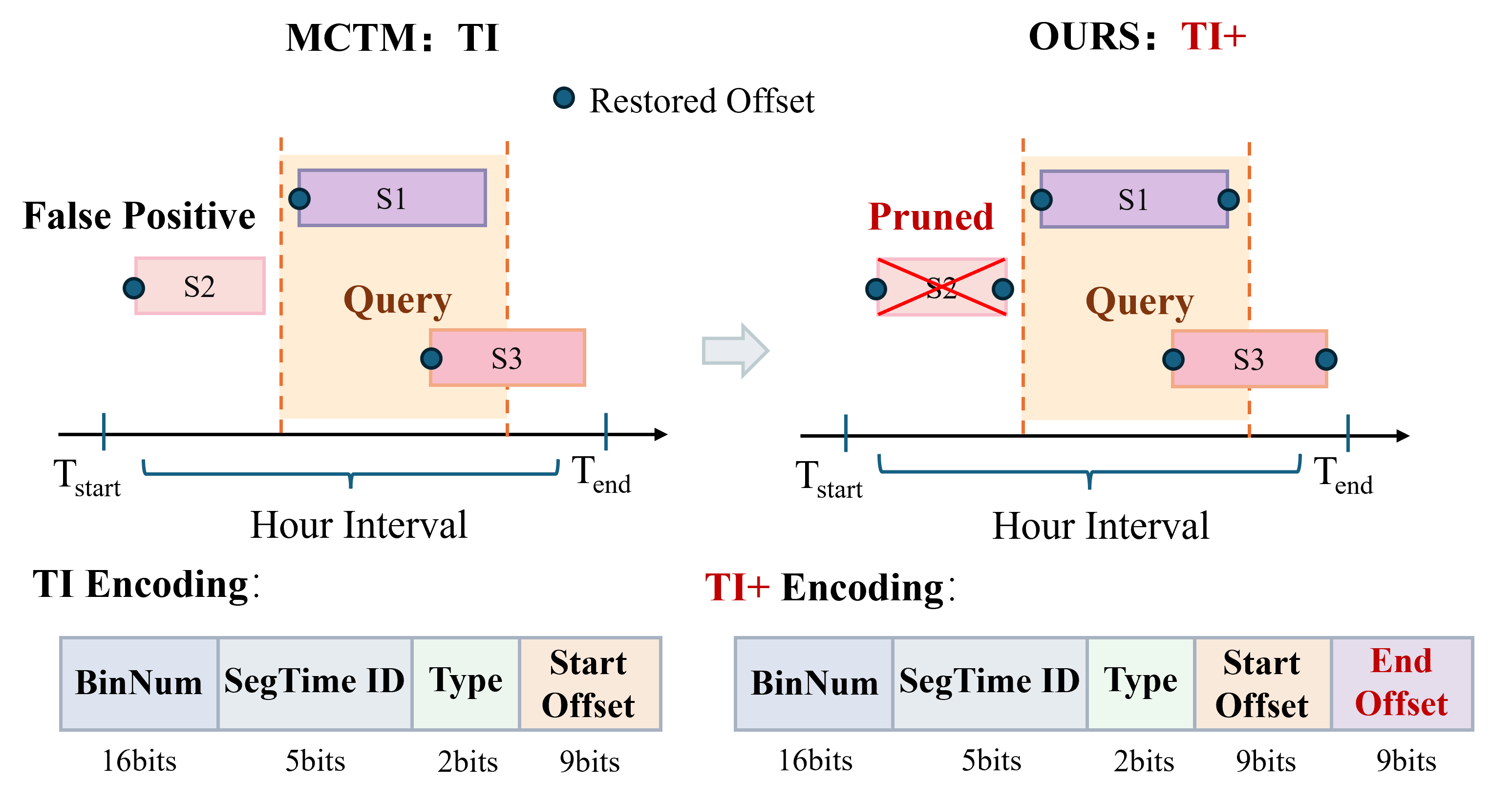}
  \caption{TI\textsuperscript{+}: type-3 entry for intra-hour boundary pruning.}
  \label{fig:ti-plus}
\end{figure}

AeroMesa introduces \textbf{type 3} for this case,
as illustrated in Figure~\ref{fig:ti-plus}.
When a segment is entirely contained within one hour interval, it is
assigned $\mathit{type}=3$ and its index entry encodes a second
9-bit \textbf{end offset} alongside the start offset:
\[
  \text{TI}^{+} =
  \underbrace{
    \underbrace{\mathit{BinNum}}_{\text{Day Bin: 16bits}}
    \;\|\;
    \underbrace{\mathit{SegTimeID}}_{\text{hour: 5bits}}
    \;\|\;
    \underbrace{\mathit{SegType}}_{\text{type: 2bits}}
    \;\|\;
    \underbrace{\mathit{Off}_9}_{\text{9bits}}
  }_{\text{original MCTM-TI: 32bits}}
  \;\|\;
  \underbrace{\mathit{Off}_9}_{\text{9bits}}
\]
Both offsets use 9-bit binary-partition encoding with
$\approx\!7\,\text{s}$ resolution, bounding the true time from both sides,
enabling intra-hour overlap pruning directly from the index key.

\subsection{3D Extension: Altitude Filtering via ZFilter}
\label{sec:zfilter}

The spatial and temporal indexes above operate in $(x, y, t)$. In the
UAV domain, altitude is an extra query dimension: flight corridors are
vertically separated by only tens of metres, yet span several
kilometres horizontally---an extreme anisotropy that renders any naive
isotropic 3D space-filling curve (e.g., XZ3 or Z3) ineffective for the
low-altitude logistics domain.

\paragraph{The aspect-ratio problem}
Standard 3D indexes map the physical domain to the unit cube
$[0,1]^3$ via isotropic normalisation, implicitly assuming that all
dimensions have comparable extents. In the UAV setting, however, the
horizontal domain spans global scales ($L_{xy}\approx40{,}000\,\text{km}$)
while the navigable vertical airspace is tightly bounded
($L_z\approx1\,\text{km}$), yielding a disparity of
$L_{xy}/L_z\approx4\times10^4$.

\begin{figure}[htbp]
  \centering
  \includegraphics[width=1.01\linewidth]{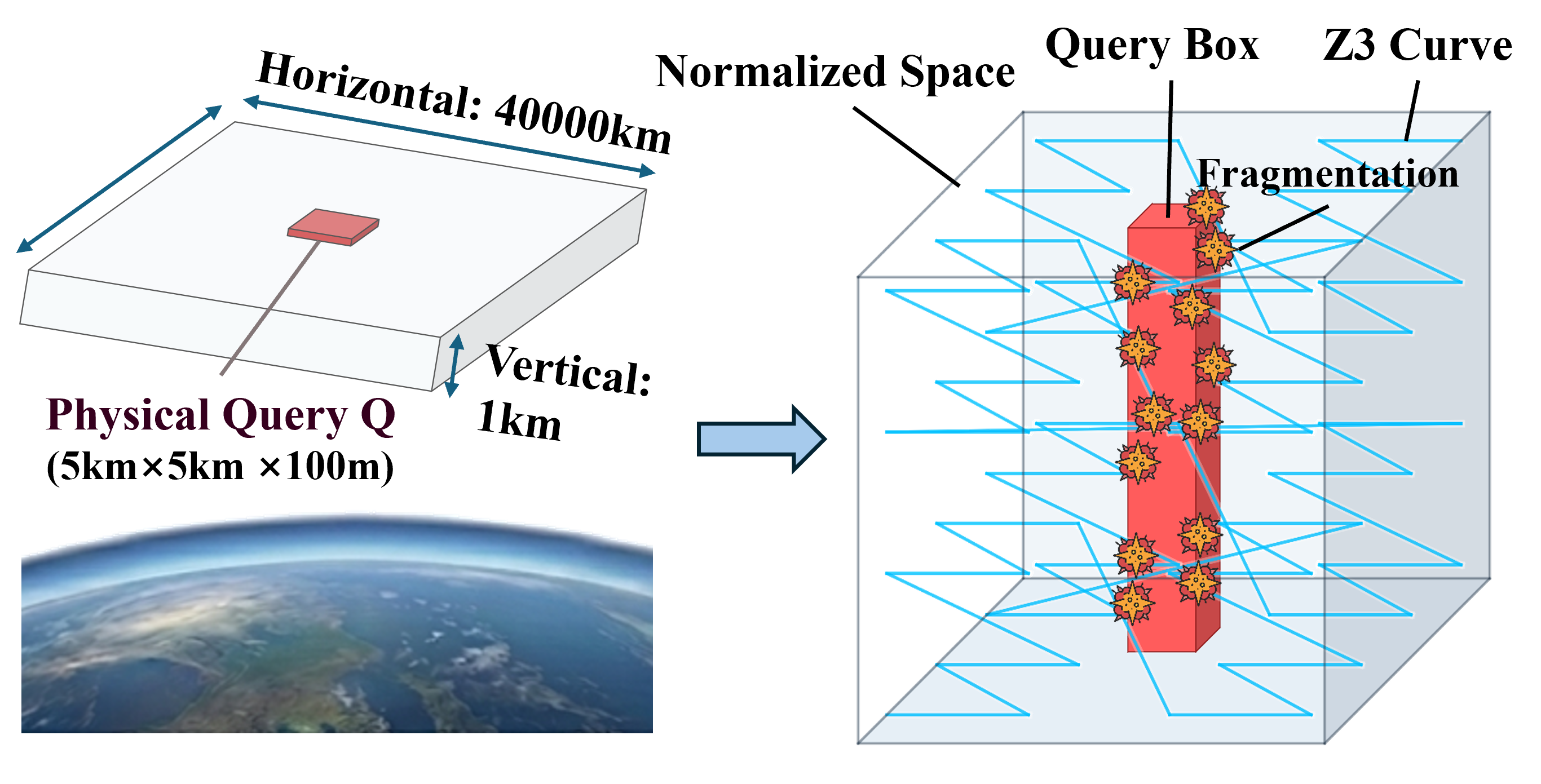}
  \caption{Geometric distortion from isotropic indexing under aspect-ratio mismatch.}
  \label{fig:isotropic-distortion}
\end{figure}

Isotropic normalisation therefore artificially stretches the
$Z$-dimension, as shown in Figure~\ref{fig:isotropic-distortion}.
A query with physical footprint $5\,\text{km} \times 100\,\text{m}$ (aspect ratio
$50:1$) becomes a query with normalised extent ratio
$\Delta q_z/\Delta q_{xy} = 800:1$. This distortion creates a dilemma
with no satisfactory resolution within the isotropic framework:

\begin{figure*}[t]
  \centering
  \includegraphics[width=\textwidth]{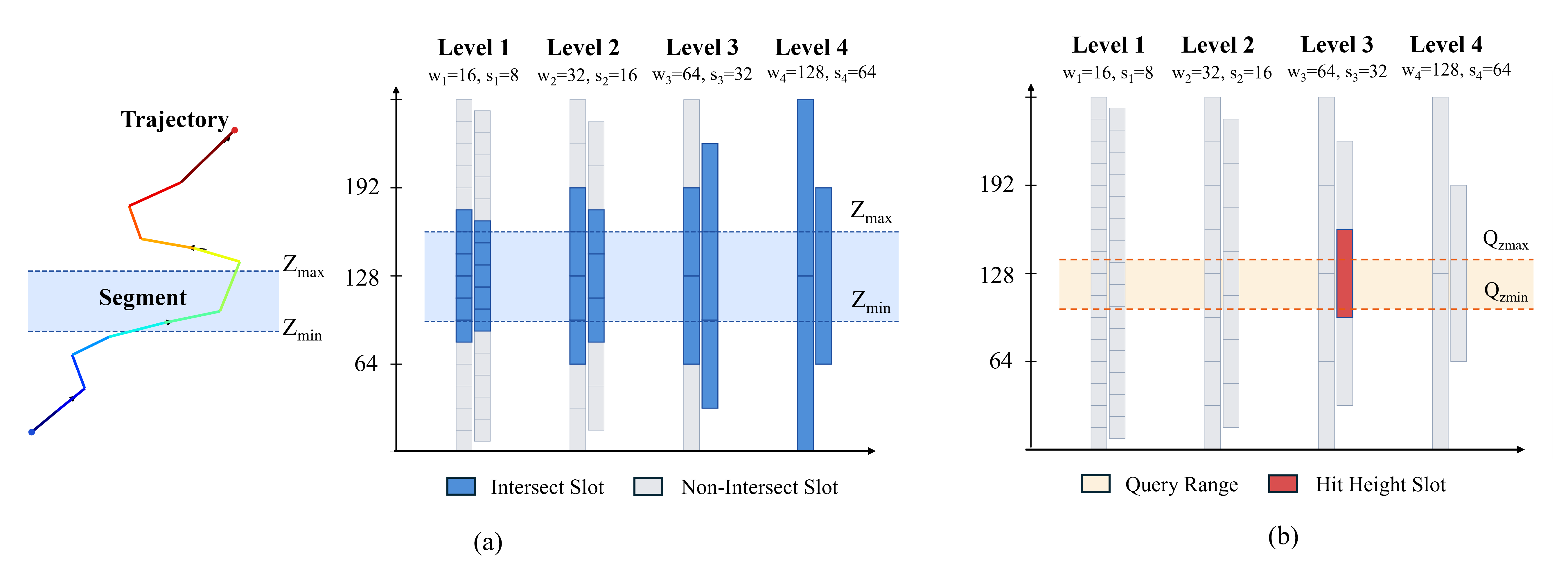}
  \caption{Multi-granularity HeightSlot: (a)~slot activation; (b)~query-time selection.}
  \label{fig:htsi}
\end{figure*}

\begin{itemize}
  \item \emph{Narrow $z$-range normalisation}: the $Z$ coordinate
    aligns with physical reality, but a $100\,\text{m}$ altitude query covers
    only $\sim\!10^{-1}$ of the normalised $Z$ axis, forcing the SFC
    to emit thousands of disjoint range intervals for even a thin
    horizontal slab (\emph{index-range fragmentation}).
  \item \emph{Wide $z$-range normalisation}: fragmentation is
    alleviated, but the same $100\,\text{m}$ altitude band maps to a negligible
    fraction of the normalised axis, making the index insensitive to
    vertical predicates and reducing it to a 2D index.
\end{itemize}

This tension is not an artifact of a particular SFC choice; it is an
intrinsic consequence of forcing a highly anisotropic domain into any
single joint spatial encoding.

\paragraph{Design response: decouple altitude from the SFC}
Rather than searching for an optimal normalisation, AeroMesa
\emph{removes altitude from the space-filling curve} and keeps the
primary ordering in $(x,y,t)$. Horizontal coordinates $(x,y)$ are
handled by Hilbert-BFS (Section~\ref{sec:spatial}), while altitude is
filtered by a dedicated \textbf{ZFilter} at the HBase RegionServer:
since each segment row already stores $[z_{\min}, z_{\max}]$ in the
\texttt{meta} column family (Section~\ref{sec:storage}),
altitude-mismatched segments are pruned \emph{before} payload transfer,
preserving horizontal index compactness and continuity.

This design is sufficient for 3D queries because the primary SI already
provides spatial selectivity: a spatial range scan yields a compact
candidate set, and ZFilter eliminates altitude mismatches within that
set entirely server-side, incurring no additional RPC round-trips.
However, a dedicated altitude index becomes necessary for 4D historical
queries: since the primary key is ordered by spatial index rather than
time, segments matching the spatial predicate are scattered across the
full temporal extent of the dataset, causing the candidate set to grow
proportionally with the queried time range and exceeding ZFilter's
pruning capacity.

\subsection{4D Extension: Height Spatio-Temporal Index (HTSI)}
\label{sec:htsi}

To efficiently resolve 4D queries, AeroMesa introduces the \textbf{Height Spatio-Temporal Index} (HTSI), a secondary structure for altitude-aware candidate generation (Figure~\ref{fig:htsi}). HTSI constrains the initial candidate set to altitude-aligned entries via a secondary index scan, while ZFilter eliminates residual false positives before payload transfer.

In this secondary access path, end-to-end performance is highly sensitive to candidate precision and scan fan-out. This raises a central design question: should altitude live in the primary SI key, or in a dedicated secondary index?

\paragraph{Why a secondary index rather than primary-key slotting?}
Embedding height-slot prefixes directly in the primary SI key fragments 2D or full-height scans into $\lceil H/\Delta h\rceil = 2^{b_z}$ disjoint sub-ranges, where $b_z = \log_2(H/\Delta h)$ is the altitude bit depth. This $O(2^{b_z})$ fragmentation penalty is fundamental to \emph{all} joint altitude-encoding strategies (e.g., an asymmetric SFC with $(a,a,c)$ bit allocation similarly yields $2^{c}$ disjoint altitude cells for a full-height query). Maintaining a separate, non-slotted primary layout to preserve 2D efficiency would require dual writes, unacceptably doubling write amplification.

AeroMesa therefore maintains a \emph{unified} primary SI for 2D/3D management and confines height-aware acceleration to HTSI. As a secondary index mapping to primary row keys, HTSI incurs acceptable write amplification. Crucially, it sidesteps the $O(2^{b_z})$ penalty entirely: for altitude-bounded queries, the multi-granularity overlapping slot design (Section~\ref{subsubsec:htsi-slots}) guarantees that any qualifying query resolves to a \emph{single} contiguous HeightSlot scan (fan-out~$= 1$). Placing HeightSlot as the leading key component after the day bin isolates altitude-bounded queries to contiguous key ranges, reducing scan fragmentation without the distortion that afflicts joint-encoding strategies.

\subsubsection{Row Key Layout}
\label{subsubsec:htsi-rowkey}

The core idea of HTSI is to exploit HBase's byte-lexicographic
ordering: by inserting a \emph{HeightSlot} field between the day-level
bin and the spatial-temporal score, all index entries sharing the same
day and altitude slot become contiguous on disk. An HTSI row key is a composite consisting of three distinct components:
\begin{equation}
  \underbrace{\mathit{BinNum}}_{16\,\text{bits}}
  \;\Big|\;
  \underbrace{\mathit{HeightSlot}}_{9\,\text{bits}}
  \;\Big|\;
  \underbrace{\mathit{LocalScore}}_{8\,\text{bytes}}
  \label{eq:htsi-rowkey}
\end{equation}
where $\mathit{BinNum} = \lfloor t_{\mathrm{unix}} / 86400 \rfloor$
is the same day bin used by the base temporal index, and
$\mathit{LocalScore}$ packs the spatial index and fine-grained
temporal fields:
\begin{equation}
  \mathit{LocalScore}
  = \bigl(\mathit{SI} \ll 5\bigr)
    \;\Big|\;
    \mathit{SegTimeID}
  \label{eq:htsi-score}
\end{equation}
where $\ll$ denotes a left bit-shift; $\mathit{SI}$ is the spatial
index score (Section~\ref{sec:spatial}) and
$\mathit{SegTimeID}\in[0,23]$ is the hourly slot encoded in 5 bits.
This ordering reflects a coarse-to-fine selectivity cascade:
$\mathit{BinNum}$ partitions the key space at day granularity first,
concentrating the scan within a bounded temporal window;
$\mathit{HeightSlot}$ then sub-partitions within that window by
altitude, yielding contiguous, doubly-constrained key ranges with
minimal scan fan-out.

\subsubsection{Multi-Granularity Height Slot Design}
\label{subsubsec:htsi-slots}

A single fixed altitude bucket size would be either too coarse, merging
corridors that queries need to distinguish, or too fine, fragmenting the
key space and inflating write amplification. Crucially, 3D queries may span
a wide range of altitude bands, so a single granularity cannot guarantee
single-slot coverage across all query types; multiple levels are therefore
necessary to match each query to the finest slot width that fully covers
its altitude span. HTSI employs \emph{four granularity levels} with window
widths $\mathbf{w} = (16, 32, 64, 128)\,\text{m}$, where the specific values
balance write amplification against selectivity: finer slots improve altitude
discrimination but multiply index entries per segment, while coarser slots
reduce amplification at the cost of pruning precision.

Within each level $k$, the altitude domain $[0, 1024)\,\text{m}$~\cite{luo2023civil}
is partitioned into overlapping windows of width $w_k$ with a 50\%
stride:
\begin{equation}
  \bigl\{[s_{k,i},\; s_{k,i}+w_k)\bigr\}_{i\ge 0},
  \quad
  s_{k,i} = i\cdot\tfrac{w_k}{2}.
  \label{eq:htsi-windows}
\end{equation}
This yields 127, 63, 31, and 15 slots for levels 1--4 respectively
(236 total, encoded in 9 bits with $2^9=512>236$). The 50\% overlap
guarantees that any altitude range of span $\Delta z \le w_k/2$ is
fully covered by one level-$k$ window, eliminating cross-slot boundary
splits for sub-granularity queries. This stride is the unique maximum
preserving the guarantee: a larger stride would introduce uncovered
gaps between adjacent windows, while a smaller stride increases write
amplification without any improvement in coverage.

As shown in Figure~\ref{fig:htsi}(a), a segment spanning
$[Z_{\min}, Z_{\max}]$ is indexed into every slot that intersects its
altitude range across all four granularity levels. Segments with
altitude outside the nominal domain are indexed only in the base TSI (the spatio-temporal index without HeightSlot prefixing),
bypassing HTSI entirely.

\paragraph{Query-time granularity selection}
Given a query altitude range $[q_z^-, q_z^+]$ with span
$\Delta q_z = q_z^+ - q_z^-$, the query engine selects the finest
level $k$ such that $\Delta q_z \le w_k/2$, ensuring the query
resolves to a single contiguous HeightSlot scan, as depicted in
Figure~\ref{fig:htsi}(b). When $\Delta q_z > 64\,\text{m}$---exceeding the
half-window of the coarsest level ($w_4/2 = 64\,\text{m}$)---no single HTSI
level can guarantee single-slot coverage, and the query falls back to
the base TSI. Queries with altitude bounds outside $[0, 1024)\,\text{m}$ are
treated as anomalous and fall back likewise.

Since UAV operations follow altitude-stratified flight corridors
governed by regulatory frameworks~\cite{luo2023civil,
kopardekar2016unmanned, bauranov2021designing}, the multi-granularity
HeightSlot layout naturally aligns with physical flight corridors,
distributing index entries across altitude levels and reducing
hot-spotting within any single granularity tier.

\paragraph{Write Amplification Analysis}
\label{subsubsec:htsi-write-amp}

HTSI incurs $\approx\!16\%$ write amplification relative to the primary
table (vs.\ $\approx\!1.2\%$ for the base TSI), as measured on our UAV
dataset (Section~\ref{subsec:dataset}), which we consider acceptable.
\section{Query Processing}
\label{sec:query}

AeroMesa supports three query types over the indexes introduced in
Section~\ref{sec:index}: temporal query, spatial query, and
spatio-temporal query. Queries translate to a few contiguous scans:
temporal query and spatio-temporal query first scan their secondary
tables to obtain candidate rowkeys, then get the matching rows from
the primary table; spatial query scans the primary table directly.
In both paths, independent scan ranges are issued to HBase
concurrently using a fixed-size thread pool, hiding per-range
round-trip latency. Predicate evaluation is pushed down to the HBase
region server via server-side filters (MBR overlap, ZFilter, and
temporal bound checks), so non-qualifying rows are discarded before
network transfer.

\subsection{Temporal Range Query}
\label{subsec:q-temporal}

A temporal range query $q = [t_{qs}, t_{qe}]$ is answered by the
TI+ index of Section~\ref{sec:ti}. Let
\begin{equation}
\left\{
\begin{aligned}
\mathrm{Bin}_s &= \lfloor t_{qs}/B \rfloor, \\
\mathrm{Bin}_e &= \lfloor t_{qe}/B \rfloor, \\
\mathrm{Seg}_s &= \lfloor (t_{qs}\bmod B)/\mathit{SegLen} \rfloor, \\
\mathrm{Seg}_e &= \lfloor (t_{qe}\bmod B)/\mathit{SegLen} \rfloor
\end{aligned}
\right.
\end{equation}
with $B = 86\,400$\,s and $\mathit{SegLen} = 3\,600$\,s,
$\mathrm{TS}_{qs}$ and $\mathrm{TS}_{qe}$ denote the
9-bit intra-segment offsets of $t_{qs}$ and $t_{qe}$. Following the segment-type discipline of MCTM,
type-0 (end) segments in $\mathrm{interval}_{qs}$ that finish before
$t_{qs}$ and type-2 (begin) segments in $\mathrm{interval}_{qe}$
that start after $t_{qe}$ must be excluded. Because the TI+ rowkey
orders $\text{type }0 < 1 < 2 < 3$ within each hour, AeroMesa
compiles the query into two contiguous index range scans that
maximise sequential I/O while minimising candidate amplification:
\begin{align}
  R_A &= \left[ \mathrm{Bin}_s \parallel \mathrm{Seg}_s
               \parallel \mathrm{type}\,0 \parallel \mathrm{TS}_{qs},
               \right. \nonumber \\
      &\hspace{3em} \left.
               \mathrm{Bin}_e \parallel \mathrm{Seg}_e
               \parallel \mathrm{type}\,2 \parallel \mathrm{TS}_{qe}
               \right], \\[2pt]
  R_B &= \left[ \mathrm{Bin}_e \parallel \mathrm{Seg}_e
               \parallel \mathrm{type}\,3 \parallel 0,
               \right. \nonumber \\
      &\hspace{3em} \left.
               \mathrm{Bin}_e \parallel \mathrm{Seg}_e
               \parallel \mathrm{type}\,3 \parallel \mathrm{TS}_{qe}
               \right].
  \label{eq:ti-window}
\end{align}
$R_A$ inherits the MCTM lower/upper sentinels and, by construction,
admits zero type-0/1/2 false positives; $R_B$ is a short auxiliary
scan that captures the single-slice (type 3) entries which $R_A$
would otherwise miss in the tail hour, while still guaranteeing
$\mathrm{startOffset}\le\mathrm{TS}_{qe}$. The middle hours
$\mathrm{Bin}_s\!:\!\mathrm{Seg}_s+1$ through
$\mathrm{Bin}_e\!:\!\mathrm{Seg}_e-1$ are fully contained inside the
query window, so all of their entries (including type 3) are admitted
by $R_A$ without further filtering.

Only the type-3 entries in the boundary hour
$(\mathrm{Bin}_s,\mathrm{Seg}_s)$ (carried by $R_A$) and, in the
single-hour case, by $R_B$ may end before $t_{qs}$; these are
validated by the dual-offset filter,
\begin{equation}
  \tau_{\mathrm{lo}}(\mathrm{startOffset}) \le \mathrm{TS}_{qe}
  \;\wedge\;
  \tau_{\mathrm{hi}}(\mathrm{endOffset})  \ge \mathrm{TS}_{qs},
\end{equation}
where $\tau_{\mathrm{lo}}$ and $\tau_{\mathrm{hi}}$ decode the lower and upper
9-bit intra-segment offsets of the entry directly from the secondary index; this check precedes any HBase payload access, so single-slice false positives never incur a payload read.
\subsection{Spatial Range Query}
\label{subsec:q-spatial}

Given a 2D box $Q_R = [x_l, x_h] \times [y_l, y_h]$, the generation of spatial
query windows proceeds in three stages:

\paragraph{Host-cell screening}
The BFS walker descends $\mathcal{T}_G$ and collects cells whose
enlarged region overlaps $Q_R$. Cells fully covered by $Q_R$ are
explicitly marked and admitted directly without descending further,
whereas partially overlapping cells are expanded to their children
until the maximum depth is reached.

\paragraph{Cache-pruned shape code resolution}
Shape code filtering is \emph{only} applied to partially overlapping
cells. For each such cell $C$ with masked address $c$, we retrieve
the set $\mathcal{P}(c)$ of occupied shape codes from the Index Cache
(Section~\ref{sec:storage}) and compute the valid overlapping subset:
\begin{equation}
  \mathcal{V}(c, Q_R)
  = \bigl\{\, p \in \mathcal{P}(c) \;\big|\;
            p \,\&\, \mathrm{ShapeCode}(Q_R \cap C) \neq 0 \,\bigr\},
\end{equation}
where $\mathrm{ShapeCode}(\cdot)$ is the bitmap representation of the
$\beta\times\beta$ sub-grid intersection.

\paragraph{Window assembly}
Each valid $(c, p)$ pair from partially overlapping cells, as well
as the inferred full-cell scan bounds from fully covered cells,
produces narrow scan ranges over the Hilbert-BFS--WAJ-linearized
local score. Two ranges are considered contiguous and merged if their
endpoints differ by at most one (e.g., $[1,5]$ and $[6,9]$ merge
into $[1,9]$); the resulting ranges are dispatched to HBase as
\texttt{Scan} commands.

\paragraph{3D spatial queries with ZFilter}
When an altitude band $[q_z^{\min}, q_z^{\max}]$ is specified, the
above 2D scan ranges remain unchanged. Instead, AeroMesa augments
the scan with a server-side HBase filter on the
\texttt{meta:zmin}/\texttt{zmax} cells
(Section~\ref{sec:zfilter}), pruning non-qualifying segments at the
region server before payload transfer.

\subsection{Spatio-Temporal Query}
\label{subsec:q-st}
A spatio-temporal query $(Q_R, [t_{qs}, t_{qe}])$ is processed by
the HTSI index of Section~\ref{sec:htsi}. 
When an altitude band is specified, its span satisfies
$\Delta q_z \le 64$\,m, and both bounds lie inside the nominal
domain $[0, 1024)$\,m, AeroMesa selects the finest HTSI level
$k^\star$ with $\Delta q_z \le w_{k^\star}/2$ and resolves the
single fully-covering slot $h^\star$
(Eq.~\eqref{eq:htsi-windows}); otherwise, the query
falls back to the height-agnostic base TSI.

The resulting rowkey range
\begin{equation}
  \bigl[\,\mathit{BinNum}\!\parallel\!h^\star\!\parallel\!\mathit{LS}_{\min},\;
        \mathit{BinNum}\!\parallel\!h^\star\!\parallel\!\mathit{LS}_{\max}\,\bigr]
\end{equation}
is contiguous on a single HBase region, where
$\mathit{LS}_{\min}$ and $\mathit{LS}_{\max}$ are the lower and
upper bounds of the concatenated field
$\mathit{LS} = \mathit{SI} \parallel \mathit{SegTimeID}$.

\section{Experimental Evaluation}
\label{sec:experiments}

This section evaluates AeroMesa across four dimensions: temporal 
range queries, 2D/3D spatial range queries, 4D spatio-temporal 
queries, and scalability.

\subsubsection{Datasets}

\label{subsec:dataset}
We evaluate the efficiency and scalability of Aeromesa using three datasets: 
\begin{itemize}
    \item \textbf{TDrive}: This dataset contains the GPS trajectories of 10,357 taxis in Beijing from February 2 to February 8, 2008. It includes approximately 15 million GPS points, with a total trajectory distance of 9 million kilometers.
    \item \textbf{Synthetic-TDrive}: For scalability evaluation, we first sample 1{,}000 trajectories from T-Drive as a 1$\times$ seed set, then generate larger datasets via random spatial translation and temporal shifting. This produces data scales up to 200$\times$ (200{,}000 trajectories).
    \item \textbf{UAV Synthetic Dataset.}
We generated 87{,}537 UAV mission trajectories over a
$10{,}000\times10{,}000$\,m area in Xuhui District, Shanghai,
using a physics-fidelity simulation pipeline combining
ROS~\cite{quigley2009ros}, Gazebo~\cite{koenig2004design},
and PX4~\cite{meier2015px4}, with the Gazebo world configured
to include realistic wind fields and weather variability.
Trajectories are planned via a grid-based A* algorithm over an
environment map derived from real OpenStreetMap road networks,
building footprints, and no-fly zone constraints, ensuring that
generated paths respect realistic urban airspace geometry.
The simulated airspace spans 0--120\,m AGL following the FAA
low-altitude operational boundary~\cite{kopardekar2016unmanned}, with
ten mission types distributed across dedicated
altitude bands under the layered airspace stratification
principle~\cite{bauranov2021designing}.
Each trajectory represents a complete end-to-end UAV mission, sequencing three primitive maneuvers---cruise, climb/descent, and hover---to replicate realistic flight dynamics.
The system boundary is provisioned up to 1024\,m to accommodate
the statutory low-altitude economy airspace ceiling~\cite{luo2023civil},
while the evaluation dataset is bounded within the regulatory 0--120\,m
AGL flight zone for lightweight UAVs; all 3D/4D query workloads execute
strictly within this dense cluster, and altitude slots above 120\,m
generate no key or metadata footprint under our storage layout.
\end{itemize}

\begin{table}[htbp]
\caption{Statistical Profiles of the UAV Dataset}
\label{tab:uav-dataset-profile}
\centering
\footnotesize
\begin{tabular}{lr}
\toprule
\textbf{Metric / Parameter} & \textbf{Value / Distribution} \\
\midrule
Total trajectories & $87{,}537$ \\
Total Points &  $37{,}741{,}478$ \\
Avg. trajectory duration & $435.43\,\text{s}$ \\
Avg. trajectory length & $3757.29\,\text{m}$ \\
Avg. spatial span (XY / Altitude) & $2851.57\,\text{m}$ / $48.71\,\text{m}$ \\
Sampling interval $\Delta t$ & $1.00\,\text{s}$ \\
Mean velocity ($v_{xy}$ / $|v_z|$) & $8.694$ / $0.218\,\text{m/s}$ \\
Mean climb / descent rate & $0.242$ / $0.224\,\text{m/s}$ \\
Maneuver ratio (Cruise : Climb : Hover) & $85.53\% : 8.31\% : 6.16\%$ \\
Corridors density (Low : Mid : High) & $42.57\% : 35.14\% : 22.29\%$ \\
\bottomrule
\multicolumn{2}{l}{\scriptsize Low/Mid/High corridors: $[0,40)$ / $[40,80)$ / $[80,120]$ m.} \\
\end{tabular}
\end{table}

\subsubsection{Settings}
\label{subsec:settings}

Each benchmark reports the average over 100 queries with distinct, randomly‑shifted spatio‑temporal windows. All experiments are repeated over 10 independent trials, and the query bounds are freshly re‑generated per run. Crucially, the relative performance gains of AeroMesa and the optimization trends in our ablation studies remain invariant under these varying conditions. All baselines are evaluated with identical concurrency and workflow configurations to ensure fairness.
We evaluated Aeromesa on hardware with 16-core CPU, 256GB RAM, and 16TB disk. The software stack includes Hadoop 3.3.4, HBase 2.5.8, and Redis 6.0.16. To align with the design 
assumptions of TMan and MCTM, AeroMesa's primary table and secondary index tables are 
deployed on the same HBase instance, while the in-memory shape-code cache 
is maintained in Redis.

\subsubsection{Baselines}

\label{subsec:baseline}
We compare AeroMesa against four representative systems.
\textbf{TMan}, the current state of the art for 2D spatial range
queries, serves as our primary spatial-index baseline; since its code
is not publicly available, we re-implemented it faithfully, and the
observed gains are consistent with those reported in the paper.
\textbf{MCTM}, which achieves state-of-the-art performance on
temporal and spatio-temporal queries, is included using the authors'
official source code.
\textbf{GeoMesa} is used as a general-purpose open-source baseline.
Finally, we include curve-only baselines---XZ2 for 2D queries and
XZ3/WXZ3 for 3D and TXZ3/TWXZ3 for 4D queries---to isolate the benefit of
higher-level index engineering over raw space-filling curves.
Among the curve-only baselines, \textbf{WXZ3} warrants particular
attention: to stress-test AeroMesa against the stronger achievable
joint-encoding strategy, we construct WXZ3 by replacing XZ3's isotropic
$2\!\times\!2\!\times\!2$ subdivision with an asymmetric
$4\!\times\!4\!\times\!2$ allocation (2/2/1 bits per $x/y/z$),
partially mitigating the aspect-ratio mismatch; any remaining gap thus
reflects an architectural limitation of joint encoding rather than a
suboptimal SFC configuration. WXZ3 is therefore included as a
\emph{constructed strong baseline}, not a contribution of this work.
All systems employing a two-level spatial encoding paradigm---including
TMan, MCTM, and AeroMesa---were configured with equivalent Redis-backed
in-memory caching for shape-code set storage. 

\subsection{Temporal Range Query}
\label{subsec:exp-temporal}

\begin{figure}[htbp]
    \centering
    \includegraphics[width=1.0\linewidth]{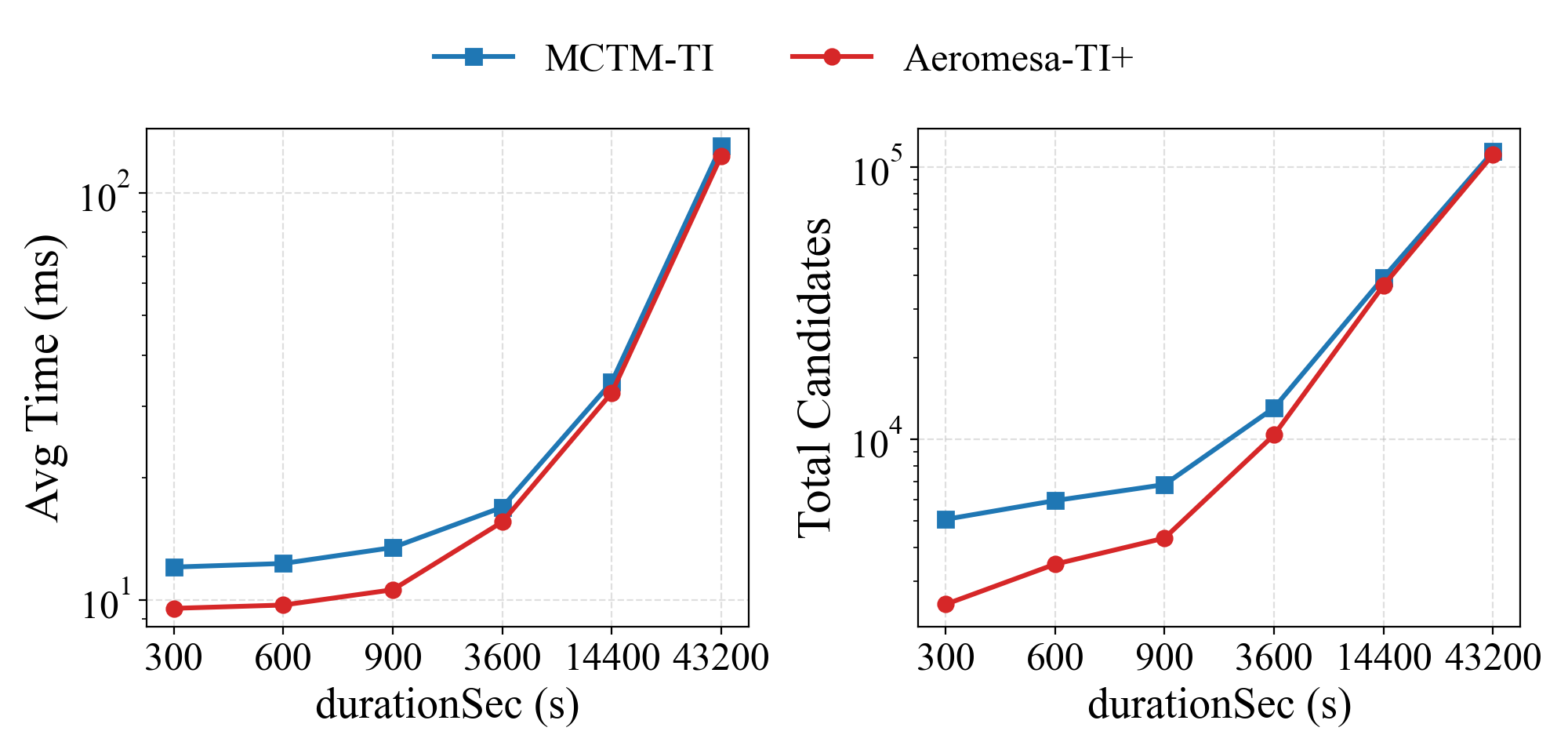}
    \caption{TI vs. TI\textsuperscript{+} on temporal range queries: average query time (left) and candidate segments (right).}
    \label{fig:exp-ti-tiplus}
\end{figure}

Figure~\ref{fig:exp-ti-tiplus} compares TI\textsuperscript{+} and TI across query durations from 300\,s to 43{,}200\,s on our UAV dataset.

TI\textsuperscript{+} consistently outperforms TI across all durations. For short windows (300--900\,s), TI\textsuperscript{+} lowers average 
query time from 12.07--13.48\,ms/query to 9.56--10.62\,ms/query 
(20.8\%--21.2\% reduction), while candidate segments decrease by 
36.4\%--51.3\%. At 3600\,s, TI\textsuperscript{+} still achieves an 
8.0\% average-time reduction (16.93 to 15.58\,ms/query) and a 20.5\% 
candidate reduction. For long windows (14{,}400--43{,}200\,s), the 
average-time gain remains 5.5\%--5.9\% (34.38 to 32.34\,ms/query and 
130.56 to 123.42\,ms/query), with 2.3\%--6.5\% fewer candidates. 
This tapering is expected: TI\textsuperscript{+} prunes false positives only at the two boundary slots, so as the window grows, the fixed pruning benefit is amortized over an increasingly larger unfiltered base from fully-contained intermediate slots.

\subsection{2D Spatial Range Query}
\label{subsec:exp-spatial-2d}

We next evaluate 2D spatial query latency on both T-Drive and UAV datasets. Query windows are grouped by side length: \textit{small} ($300\times300$\,m), \textit{medium} ($500\times500$\,m), \textit{large} ($1000\times1000$\,m), \textit{xlarge} ($2000\times2000$\,m), and \textit{xxlarge} ($5000\times5000$\,m).

\begin{figure}[htbp]
    \centering
    \includegraphics[width=1.0\linewidth]{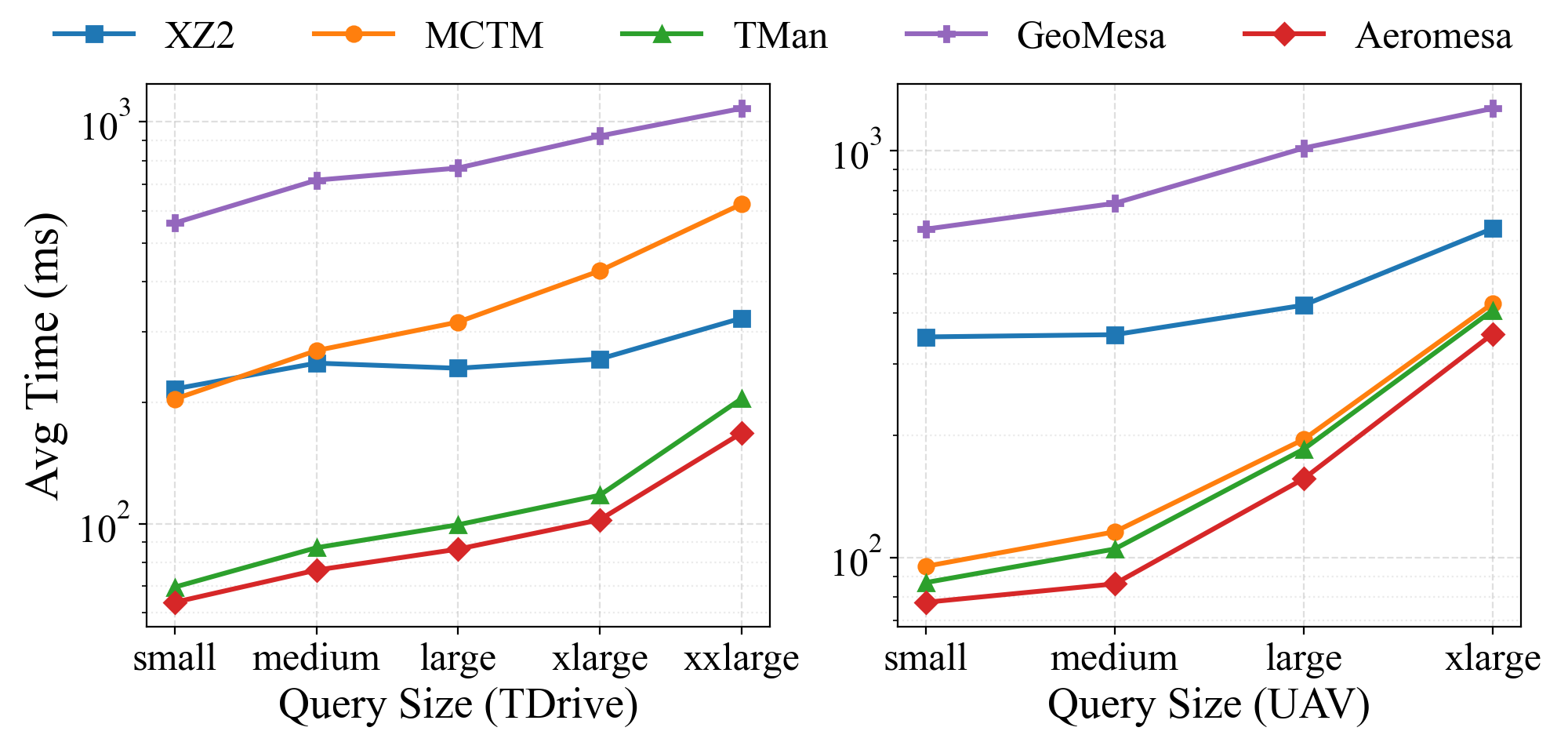}
    \caption{2D spatial query latency (left: T-Drive, right: UAV).}
    \label{fig:exp-spatial-main}
\end{figure}

Figure~\ref{fig:exp-spatial-main} shows AeroMesa achieves the lowest latency on both datasets. On T-Drive it reduces latency by 8.3\%--17.9\% over TMan, 68.8\%--76.0\% over MCTM, 48.2\%--70.6\% over XZ2, and 84.4\%--89.3\% over GeoMesa; results on UAV are consistent (10.6\%--17.9\% over TMan; 72.1\%--88.4\% over GeoMesa).

\subsubsection{Ablation: Independent Gains of Hilbert-BFS and WAJ}
\label{subsec:exp-spatial-ablation}

We isolate each component by comparing: \textit{TMan} (XZ2+Jaccard), \textit{XZ2+WAJ}, \textit{Hilbert-BFS+Jaccard}, and \textit{AeroMesa} (Hilbert-BFS+WAJ).

\begin{figure}[htbp]
    \centering
    \includegraphics[width=1.0\linewidth]{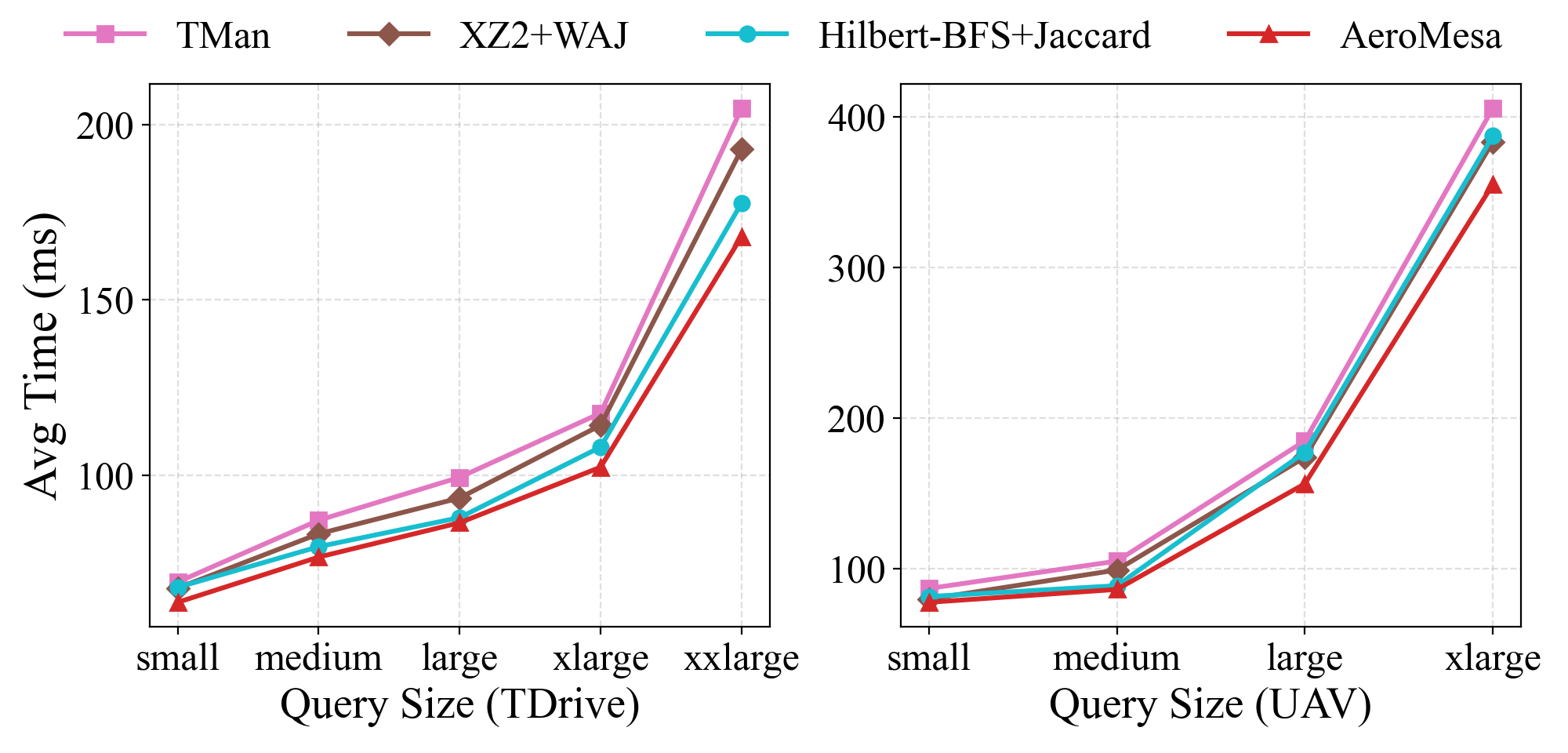}
    \caption{Ablation on 2D spatial queries (left: T-Drive, right: UAV).}
    \label{fig:exp-spatial-ablation}
\end{figure}

Figure~\ref{fig:exp-spatial-ablation} shows both components contribute independently: WAJ alone reduces latency by 2.5\%--5.9\% (T-Drive) and 5.5\%--8.1\% (UAV); Hilbert-BFS alone by 2.2\%--13.2\% and 4.1\%--15.5\%. Combined, AeroMesa improves 8.3\%--17.9\% over TMan (T-Drive) and 10.6\%--17.9\% (UAV), with WAJ adding a further 1.7\%--6.3\% and 2.8\%--11.8\% on top of Hilbert-BFS alone.

\subsection{3D Spatial Range Query}
\label{subsec:exp-spatial-3d}

%
%
We use default \textit{XZ3} (altitude $\in[0,1000]$\,m) as the primary
curve baseline, with query footprints of $1000\times1000$\,m and
altitude-range classes $\{10,25,50,100,250\}$\,m. As described in
Section~\ref{subsec:baseline}, \textit{WXZ3} is the constructed
strong-baseline counterpart of XZ3, replacing the isotropic
$2\!\times\!2\!\times\!2$ subdivision (1/1/1 bits per $x/y/z$) with
$4\!\times\!4\!\times\!2$ (2/2/1 bits) to form a standard 32-ary tree.
For example, after 9 levels, its 18-bit horizontal
($\Delta_{xy} \approx 152\,\text{m}$) and 9-bit vertical
($\Delta_z = 2\,\text{m}$) resolutions tightly map a
$1000\!\times\!1000\!\times\!10\,\text{m}$ query to a compact
$6\!\times\!6\!\times\!5$ cell block, making the index granularity
closely fit our segments and query box.
%
%

\begin{figure}[htbp]
    \centering
    \includegraphics[width=1.0\linewidth]{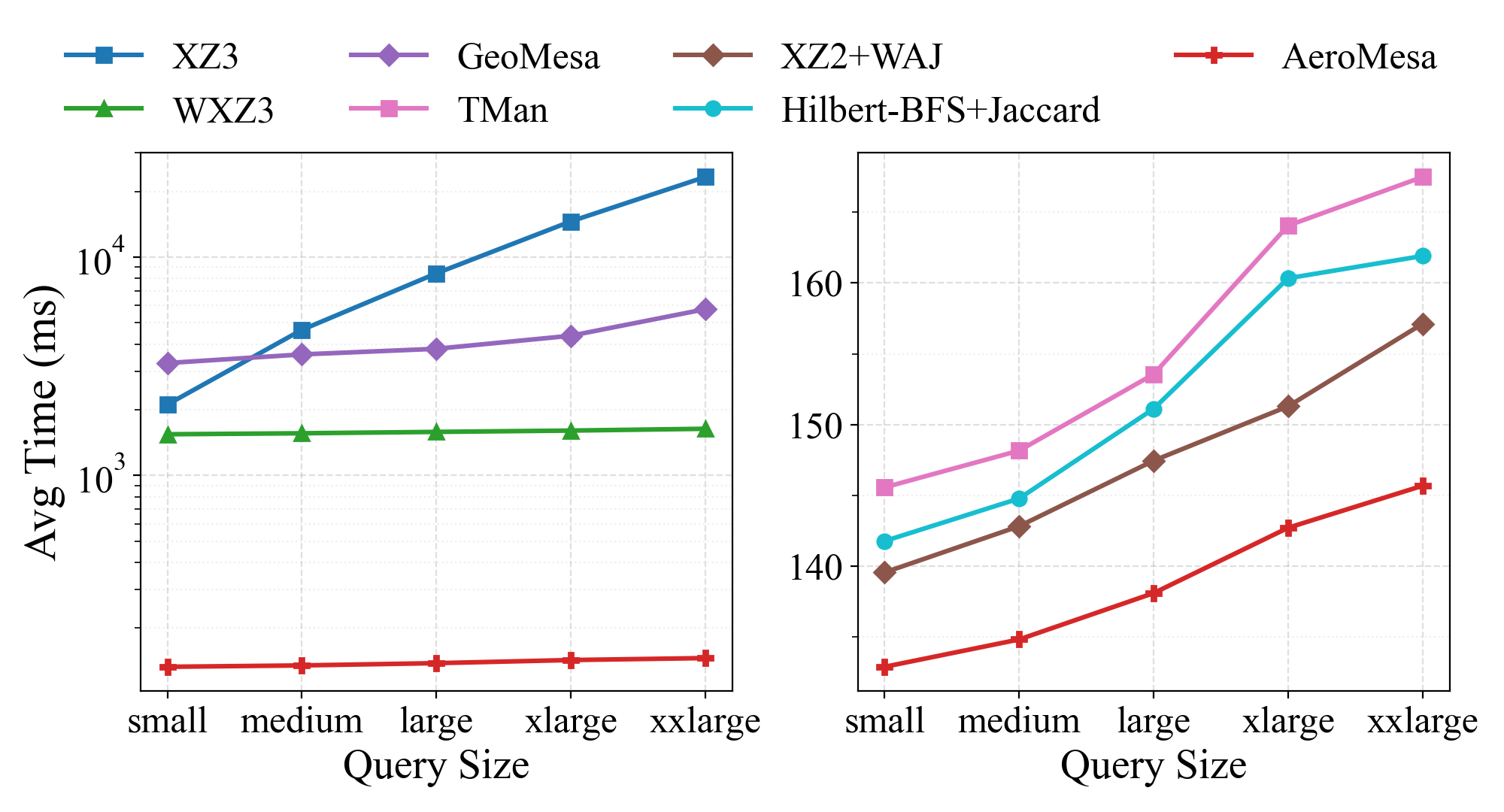}
    \caption{3D spatial query latency (left: main baseline comparison; right: ablation of Hilbert-BFS and WAJ).}
    \label{fig:exp-spatial-3d}
\end{figure}

Figure~\ref{fig:exp-spatial-3d} (left) shows AeroMesa remains nearly flat at 132.9--145.7\,ms ($1.10\times$), while XZ3 escalates $11.03\times$ (to 23293.2\,ms). WXZ3 stays much flatter ($1.06\times$, 1543.9--1635.9\,ms) but is still consistently slower than AeroMesa. AeroMesa reduces latency by 93.7\%--99.4\% over XZ3, 91.1\%--91.4\% over WXZ3, and 95.9\%--97.5\% over GeoMesa. The right panel confirms the ablation trend: WAJ contributes 3.6\%--7.8\%, Hilbert-BFS 1.6\%--3.3\%, and combined they yield 8.7\%--13.0\% improvement over the TMan baseline, with WAJ adding 6.3\%--11.0\% on top of Hilbert-BFS.

\subsection{Spatio-Temporal Query (4D)}
\label{subsec:exp-spatiotemporal}

We evaluate 4D queries in two complementary settings. In all queries, the query horizontal footprint is fixed to $1000 \times 1000$\,m. First, we fix the temporal window at $30000$\,s and vary altitude range over $\{5,15,30,60,120,240\}$\,m. Second, we fix height range at $15$\,m and vary duration over $\{3600,7200,15000,30000,60000,180000\}$\,s.

TXZ3 uses default XZ3 as the spatial index key, while TWXZ3 uses 
weighted XZ3; \textbf{base TSI} and \textbf{AeroMesa-HTSI} correspond 
to the spatio-temporal index without and with the HeightSlot prefix, 
respectively (see Section~\ref{sec:htsi}).

\begin{figure}[htbp]
    \centering
    \includegraphics[width=1.0\linewidth]{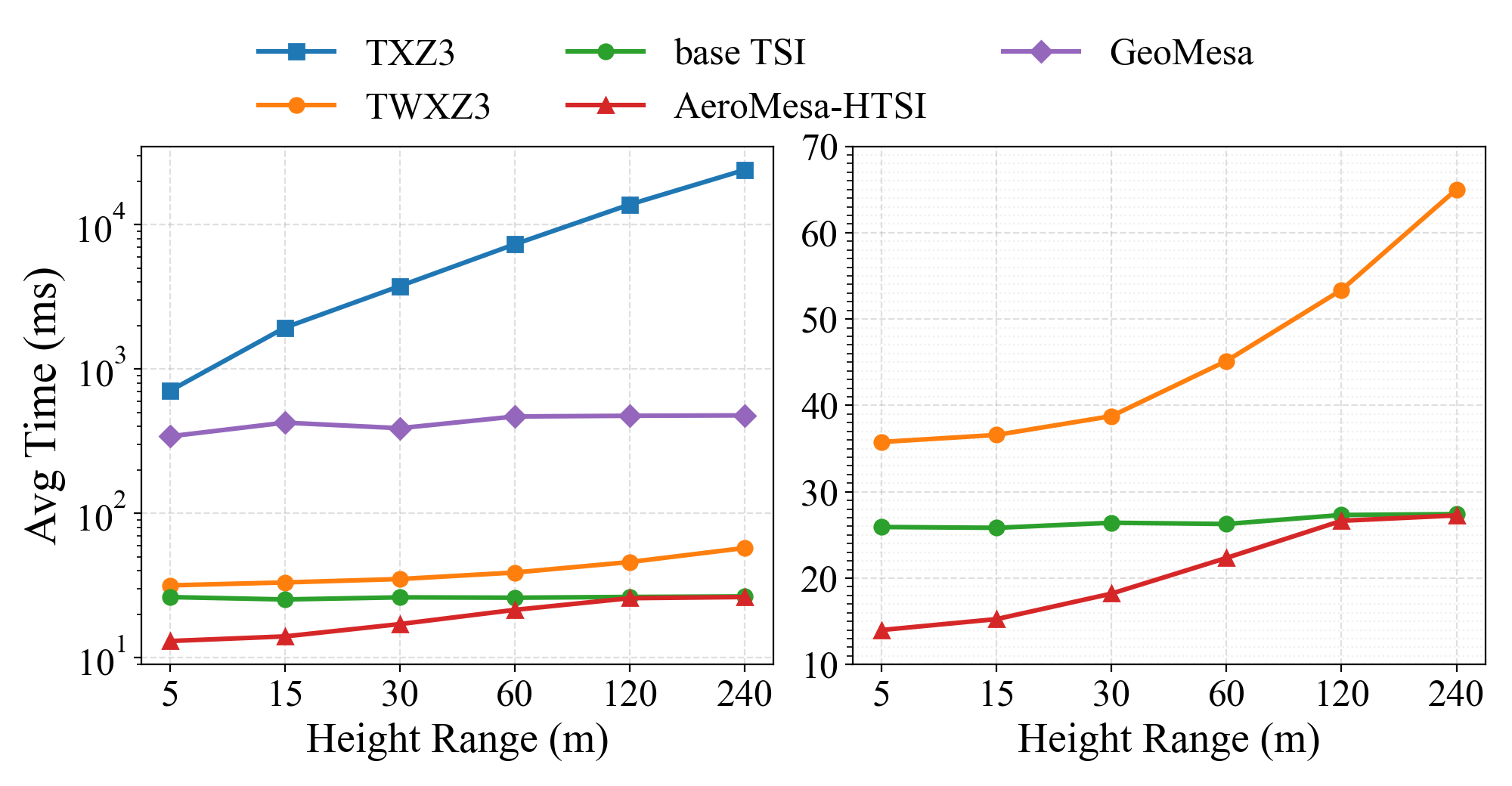}
    \caption{4D spatio-temporal query with fixed time (left: main comparison; right: TWXZ3, base TSI, and HTSI).}
    \label{fig:exp-st-height}
\end{figure}

\begin{figure}[htbp]
    \centering
    \includegraphics[width=1.0\linewidth]{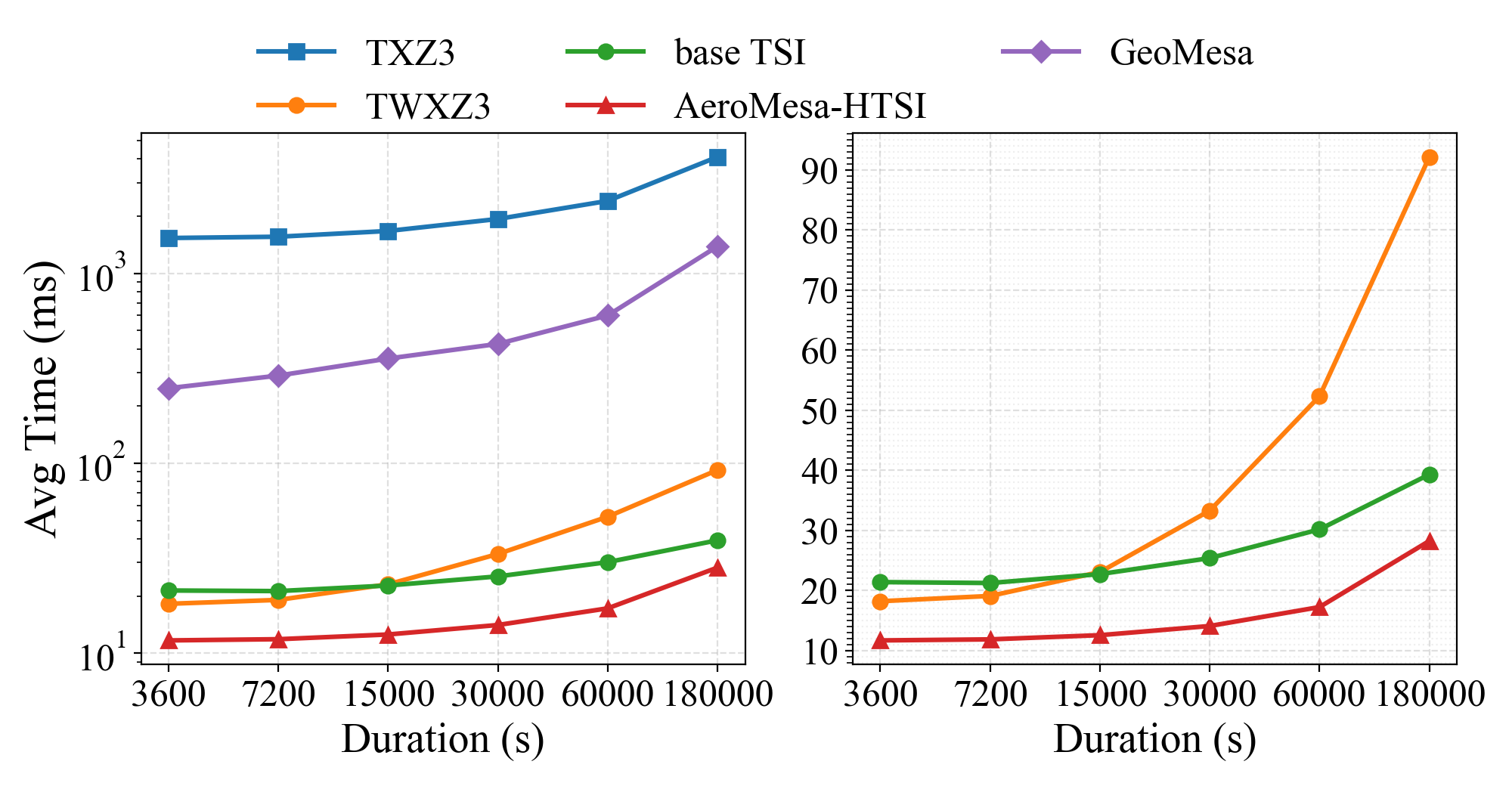}
    \caption{4D spatio-temporal query with fixed height (left: main comparison; right: TWXZ3, base TSI, and HTSI).}
    \label{fig:exp-st-time}
\end{figure}

Figure~\ref{fig:exp-st-height} shows that, under a fixed $30000$\,s window, HTSI grows from 13.07\,ms to 26.37\,ms as altitude range increases, versus 25.37--26.63\,ms for base TSI, yielding 1.0\%--50.4\% lower latency. Relative to TXZ3, TWXZ3, and GeoMesa, HTSI reduces latency by 98.1\%--99.9\%, 43.8\%--58.9\%, and 94.5\%--96.7\%, respectively. The gain over base TSI comes from HeightSlot pruning: candidate segments drop from 1603.64 to 418.54--1232.65 while the merged scan range count remains unchanged for both variants. The larger gap against TXZ3 is due to joint-encoding fragmentation: average merged ranges rise to 15{,}736--554{,}328 for TXZ3, compared with 105.65--491.84 for TWXZ3 and 57.44 for base TSI/HTSI.

Figure~\ref{fig:exp-st-time} shows the same pattern when height is fixed at $15$\,m and duration increases. HTSI rises from 11.67\,ms to 28.24\,ms, compared with 21.23\,ms to 39.31\,ms for base TSI, preserving a 28.1\%--45.4\% latency reduction across all durations. Against TXZ3, TWXZ3, and GeoMesa, HTSI remains 99.2\%--99.3\%, 35.9\%--69.4\%, and 95.3\%--97.9\% faster, respectively. Again, the HTSI--base TSI gap comes from lower candidate counts (386.76--1470.66 vs.\ 1197.63--3825.66), while both variants keep the same merged scan ranges (43.9--128.95); in contrast, TXZ3 still incurs much heavier fragmentation, with average merged ranges reaching 35{,}665--104{,}875, while TWXZ3 reduces this to 98.56--290.08.

\subsection{Scalability}
\label{subsec:exp-scalability}

We evaluate scalability on Synthetic-TDrive with data scales from 1$\times$ to 200$\times$, using 2D query workload ($500\times500$\,m).

\begin{figure}[htbp]
    \centering
    \includegraphics[width=1.0\linewidth]{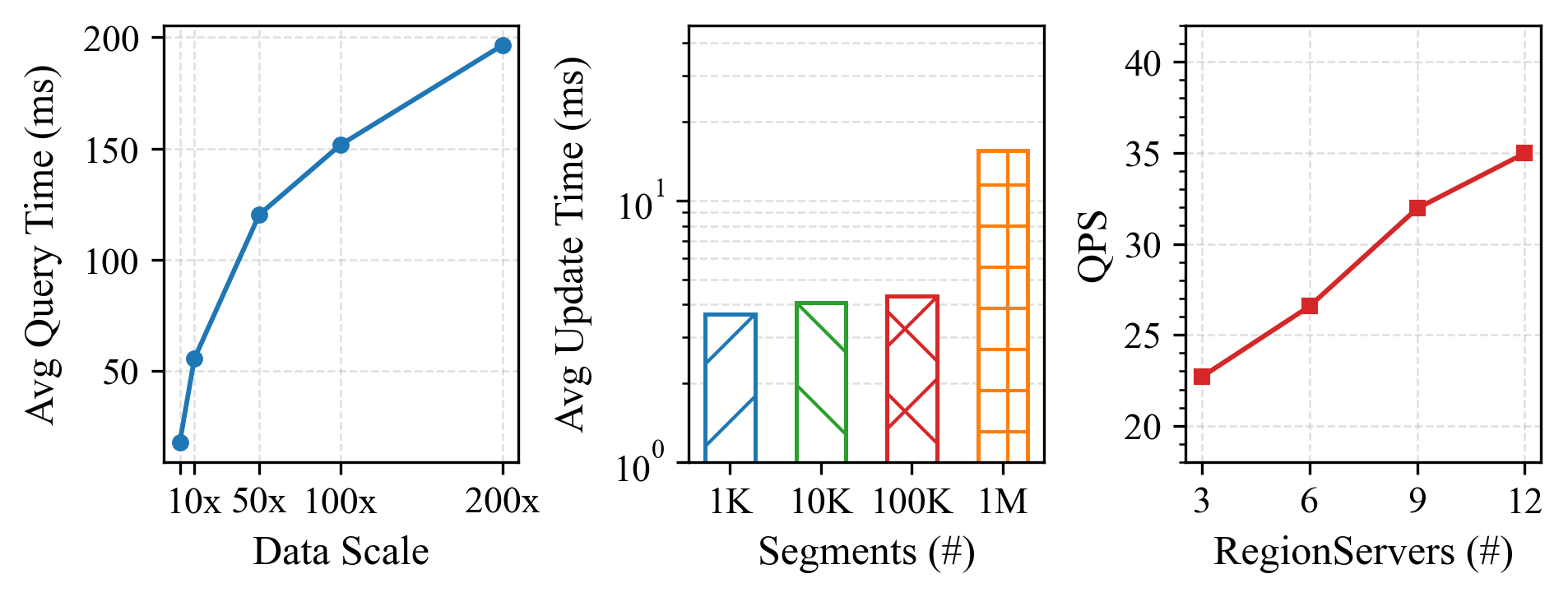}
    \caption{Scalability of AeroMesa 
    }
    \label{fig:exp-scalability}
\end{figure}

Figure~\ref{fig:exp-scalability} (left) shows query latency grows from 17.82\,ms ($1\times$) to 196.47\,ms ($200\times$)---an $11.03\times$ increase for $200\times$ data growth, confirming sub-linear read scalability for AeroMesa.

The middle panel reports the amortized write cost of batch-inserting 4{,}000 segments into a store of varying size. The per-segment update time stays stable at 3.65--4.32\,ms for up to 100{,}000 existing segments, and rises moderately to 15.53\,ms at 1{,}000{,}000 segments---a $4.3\times$ increase for a $1000\times$ growth in store size. This sub-linear growth is attributed to the main+delta LSM design (Section~\ref{sec:update}), which amortizes WAJ re-encoding across background compaction batches.

The right panel reports distributed query throughput under the same intensive workload comprising 1{,}000 simultaneous queries (each with a $500 \times 500$ meter 2D spatial window) on the T-Drive dataset. As RegionServer instances increase from 3 to 12, throughput rises from 22.7 to 35.0 QPS, a 54.1\% improvement. The gain is sub-linear, attributed to resource contention and RPC overhead under high concurrency.
\section{Conclusion}
\label{sec:conclusion}
This paper presents AeroMesa, an efficient data management system
for multi-dimensional spatio-temporal trajectories built on Apache
HBase and Redis.
By decoupling horizontal indexing from altitude-aware secondary
indexing, AeroMesa eliminates the row-key interval fragmentation
inherent in joint spatial encodings while preserving full support
for $(x,y)$, $(x,y,t)$, $(x,y,z)$, and $(x,y,z,t)$ queries within
a unified storage framework.
Extensive evaluations demonstrate that AeroMesa consistently
outperforms all baselines: 3D/4D query latency is reduced by up to
$30\times$ over XZ3/TXZ3, 2D spatial latency by up to 17.9\% over
TMan, and temporal candidates by up to 51.3\% over MCTM, with
sub-linear scalability confirmed under $200\times$ data expansion.
Interesting future work includes:
1)~scaling \textsc{AeroMesa} to larger distributed deployments by
devising workload-aware partitioning strategies to mitigate
high-concurrency RPC overhead and balance storage locality; and
2)~extending \textsc{AeroMesa} to support efficient $k$-nearest-neighbor
and trajectory similarity queries.

\bibliographystyle{IEEEtran}
\bibliography{sample-base} 

\end{document}